\documentclass[a4paper,11pt]{article}

\pdfoutput=1

\usepackage{jcappub_ver}
\usepackage{amsmath}
\usepackage{graphicx}

\usepackage{physics}
\usepackage{pgfplots}

\usepackage[english]{babel}
\usepackage[utf8]{inputenc}
\usepackage{pdflscape}
\usepackage{enumerate}
\usepackage{amsbsy}
\usepackage{amsmath} %used by default
\usepackage{graphics}
\usepackage{mathrsfs}
\usepackage{wrapfig}
\usepackage{mathtools}

\usepackage{amsfonts}
\usepackage{pstricks}
\usepackage{color}
\usepackage{setspace}
\usepackage{tensor}
\usepackage[ruled,vlined]{algorithm2e}
\usepackage{array}
\usepackage{booktabs}
\usepackage{footmisc}

\usepackage[normalem]{ulem}
\usepackage{float}

\newcommand{\bea}{\begin{eqnarray}} \newcommand{\eea}{\end{eqnarray}}
\newcommand{\el}{\nonumber \\}
\newcommand{\re}[1]{(\ref{#1})}

\newcommand{\pat}{\partial}
\renewcommand{\sec}[1]{section \ref{#1}}
\newcommand{\fig}[1]{figure \ref{#1}}
\newcommand{\tab}[1]{table \ref{#1}}

\newcommand{\para}{\paragraph}

\renewcommand*{\vec}[1]{\mathbf{#1}}
\newcommand{\PR}{\mathcal{P}_\mathcal{R}}
\newcommand{\R}{\mathcal{R}}
\newcommand{\mr}[1]{\mathrm{#1}}

\renewcommand{\a}{\alpha}
\renewcommand{\b}{\beta}

\renewcommand{\d}{\delta}

\newcommand{\ha}{\frac{1}{2}}
\newcommand{\rmd}{\mathrm{d}}

\newcommand{\sign}{\mathrm{sign}}

\newcommand{\ie}{i.e.\ }
\newcommand{\eg}{e.g.\ }

\newcommand{\adot}{\dot{a}}

\newcommand{\Mpl}{M_{{}_{\mathrm{Pl}}}}

\newcommand{\de}{\mathrm{d}}

\newcommand{\phil}{\bar\phi}
\newcommand{\pil}{\bar\pi}
\newcommand{\kpbh}{k_\mr{PBH}}

\newcommand{\phif}{{\bar\phi}_\mr{f}}
\newcommand{\phii}{{\bar\phi}_\mr{i}}
\newcommand{\kc}{k_\mathrm{c}}

\newcommand{\paperI}{$Paper\,I$}

\begin{document}

\title{Implications of stochastic effects for primordial black hole production in ultra-slow-roll inflation}

\author[a]{Daniel G. Figueroa}
\author[b]{\hspace*{-2mm}, Sami Raatikainen}
\author[b]{\hspace*{-2mm}, Syksy R\"{a}s\"{a}nen}
\author[c]{\hspace*{-2mm}, and Eemeli Tomberg}

\affiliation[a]{Instituto de F\'isica Corpuscular (IFIC), Consejo Superior de Investigaciones Cient\'ificas (CSIC) and Universitat de Valencia, E-46980, Valencia, Spain}

\affiliation[b]{University of Helsinki, Department of Physics and Helsinki Institute of Physics,\\ P.O. Box 64, FIN-00014 University of Helsinki, Finland}

\affiliation[c]{Laboratory of High Energy and Computational Physics, National Institute of Chemical
Physics and Biophysics, R\"{a}vala pst. 10, 10143 Tallinn, Estonia}

% e-mail addresses: one for each author, in the same order as the authors
\emailAdd{daniel.figueoa@ific.uv.es}
\emailAdd{sami.raatikainen@helsinki.fi}
\emailAdd{syksy.rasanen@iki.fi}
\emailAdd{eemeli.tomberg@kbfi.ee}

\begin{flushleft}
	\hfill		 HIP-2021-31/TH \\
\end{flushleft}

\abstract{We study the impact of stochastic noise on the generation of primordial black hole (PBH) seeds in ultra-slow-roll (USR) inflation with numerical simulations. We consider the non-linearity of the system by consistently taking into account the noise dependence on the inflaton perturbations, while evolving the perturbations on the coarse-grained background affected by the noise. We capture in this way the non-Markovian nature of the dynamics, and demonstrate that non-Markovian effects are subleading. Using the $\Delta N$ formalism, we find the probability distribution $P(\mathcal{R})$ of the comoving curvature perturbation $\mathcal{R}$. We consider inflationary potentials that fit the CMB and lead to PBH dark matter with $i)$ asteroid, $ii)$ solar, or $iii)$ Planck mass, as well as $iv)$ PBHs that form the seeds of supermassive black holes. We find that stochastic effects enhance the PBH abundance by a factor of $\mathcal{O}(10)-\mathcal{O}(10^8)$, depending on the PBH mass. We also show that the usual approximation, where stochastic kicks depend only on the Hubble rate, either underestimates or overestimates the abundance by orders of magnitude, depending on the potential. We evaluate the gauge dependence of the results, discuss the quantum-to-classical transition, and highlight open issues of the application of the stochastic formalism to USR inflation.}

\maketitle
  
\setcounter{tocdepth}{2}

\setcounter{secnumdepth}{3}

\section{Introduction} \label{sec:intro}

\para{Stochastic ultra-slow-roll inflation and primordial black holes.}

Inflation is the most successful scenario to explain the primordial universe~\cite{Akrami:2018odb}, predicting that the average spatial curvature is negligible, and that the primordial perturbations observed in the cosmic microwave background (CMB) and large scale matter perturbations are almost scale-invariant, highly Gaussian, mostly adiabatic, dominated by a growing scalar mode, and statistically very homogeneous and isotropic \cite{Starobinsky:1979ty, Starobinsky:1980te, Kazanas:1980tx, Guth:1981, Sato:1981, Mukhanov:1981xt, Linde:1981mu, Hawking:1981fz, Albrecht:1982wi, Chibisov:1982nx, Hawking:1982cz, Guth:1982ec, Starobinsky:1982ee, Sasaki:1986hm, Mukhanov:1988jd}. In the simplest inflationary models, a scalar field called the inflaton rolls down its potential slowly, with Hubble friction and potential push balanced, so that its acceleration is negligible, like for a falling parachute. This is called slow-roll (SR).

If the potential has a very flat section or a shallow minimum, the potential push is instead negligible, and the inflaton acceleration is dominated by friction, leading to an exponentially falling velocity, like for a brick sliding on a flat surface. Such a regime is called ultra-slow-roll (USR). Perturbations generated in USR are far from scale-invariant, and can be highly non-Gaussian. Therefore, the inflaton cannot be in USR when the modes observed in the CMB are generated. However, as the inflaton perturbations are enhanced by the fact that its background velocity is small, USR outside the CMB region could generate large perturbations that seed primordial black holes (PBH) \cite{Garcia-Bellido:2017mdw, Ezquiaga:2017fvi, Kannike:2017bxn, Germani:2017bcs, Motohashi:2017kbs, Gong:2017qlj, Ballesteros:2017fsr, Hertzberg:2017dkh, Biagetti:2018pjj, Ezquiaga:2018gbw, Rasanen:2018fom, Drees:2019xpp, Fu:2020lob, Ballesteros:2020sre, Gangopadhyay:2021kmf, Wang:2021kbh}. PBHs are a longstanding dark matter candidate \cite{Chapline:1975, Dolgov:1992pu, Ivanov:1994pa, Yokoyama:1995ex, GarciaBellido:1996qt, Jedamzik:1996mr, Ivanov:1997ia, Blais:2002nd}, which can constitute all of the dark matter, or they can act as seeds for supermassive black holes \cite{Carr:2020gox, Carr:2020xqk, Green:2020jor}. Large inhomogeneities on small scales could also affect big bang nucleosynthesis \cite{Arbey:2020} and recombination \cite{Jedamzik:2020, Rashkovetskyi:2021rwg}.

During inflation, perturbations are constantly created on sub-Hubble scales. Due to accelerated expansion, perturbations are stretched beyond the Hubble radius, creating a phase space populated by modes with exponentially different wavelengths. In the leading approximation, perturbations evolve according to linear perturbation equations around a Friedmann--Lema\^itre--Robertson--Walker (FLRW) background \cite{Salopek:1992qy, Wands:2000dp}. Beyond the leading approximation, modes with different wavelengths couple. Small wavelength modes can be treated as an environment for the system of large wavelength modes. On super-Hubble scales, modes are squeezed and their probability distribution classicalises~\cite{Habib:1992ci, Polarski:1995jg, Calzetta:1995ys, Bellini:1996uh, Lesgourgues:1996jc, Kiefer:1998qe, Kiefer:1998jk, Casini:1998wr, Kiefer:2008ku}. Once the originally quantum small wavelength modes become super-Hubble, they can be coarse-grained and treated as stochastic noise for the effectively classical long wavelength modes. This effect is captured by the formalism of stochastic inflation, first introduced in~\cite{Starobinsky:1986} and further developed in~\cite{Morikawa:1989xz, Salopek:1990jq, Salopek:1990re, Habib:1992ci, Mijic:1994vv, Starobinsky:1994, Bellini:1996uh, Tsamis:2005hd, Woodard:2005cv, Martin:2005hb, vanderMeulen:2007ah, Finelli:2008zg, Beneke:2012kn, Levasseur:2013ffa, Gautier:2013aoa, Levasseur:2013tja, Fujita:2013cna, Garbrecht:2013, Fujita:2014tja, Levasseur:2014ska, Garbrecht:2014dca, Onemli:2015, Vennin:2015hra, Boyanovsky:2015tba, Boyanovsky:2015jen, Burgess:2015ajz, Moss:2016uix, Grain:2017dqa, Collins:2017haz, Pattison:2017mbe, Tokuda:2017fdh, Prokopec:2017vxx, Cruces:2018cvq, Firouzjahi:2018vet, Markkanen:2019, Pattison:2019hef, Prokopec:2019srf, Ezquiaga:2019ftu, Pinol:2020cdp, Firouzjahi:2020jrj, De:2020hdo, Ando:2020fjm, Figueroa:2020jkf, Pattison:2021oen, Rigopoulos:2021nhv, Cruces:2021iwq}. 

In typical SR inflationary scenarios stochastic effects are small, though there can be exceptions, especially when inflation lasts much longer than 60 e-folds~\cite{Tsamis:2005hd, Woodard:2005cv, Ando:2020fjm}. However, in USR stochastic effects can be expected to be important. This is due to three reasons. First, as the potential push is absent and the field velocity decays exponentially, the field evolution is more sensitive to stochastic kicks, especially if USR lasts long. Second, the USR regime can produce larger perturbations than standard SR, and hence give stronger stochastic kicks. Third, the main observable related to PBH abundance -- the prevalence of high density regions -- is exponentially sensitive to the amplitude of the tail of the curvature perturbation distribution, where the usual Gaussian contribution is highly suppressed and the strong non-Gaussianities from USR may be important\footnote{As pointed out in \cite{Ando:2020fjm}, stochastic effects can also have a large effect on the CMB even when the USR region of the potential is far from the CMB region, because some rare modes can remain in the part of the potential where stochastic effects are important for so long that their wavelength stretches to CMB scales.}.

The effect of stochastic kicks on the power spectrum in USR has been studied in \cite{Vennin:2015hra, Biagetti:2018pjj, Ezquiaga:2018gbw, Cruces:2018cvq, Firouzjahi:2018vet, Pattison:2019hef, Ballesteros:2020sre, De:2020hdo, Ando:2020fjm, Firouzjahi:2020jrj,Cruces:2021iwq, Rigopoulos:2021nhv}, see also~\cite{Vennin:2015hra, Firouzjahi:2018vet,  Firouzjahi:2020jrj} for higher moments. Using the $\Delta N$ formalism \cite{Sasaki:1995aw, Sasaki:1998ug, Wands:2000dp, Lyth:2004gb} it has been demonstrated that stochastic effects can produce an exponential tail in the probability distribution $P(\mathcal{R})$ of the curvature perturbation $\mathcal{R}$, which dominates over the Gaussian tail expected from linear theory \cite{Pattison:2017mbe, Ezquiaga:2019ftu}. As PBHs are produced from rare perturbations, the amplitude of the power spectrum may then not be enough to determine the PBH abundance, even when the power spectrum correctly describes perturbations close to the mean.

The calculations in \cite{Pattison:2017mbe, Ezquiaga:2019ftu} were in the context of SR, and in our recent Letter~\cite{Figueroa:2020jkf} -- which we refer to as \paperI~-- we presented the first stochastic calculation in USR inflation. We calculated numerically the probability distribution $P(N)$ for the number of e-folds $N$ of inflation in the $\Delta N$ formalism, which gives $P(\mathcal{R})$. We simulated a large number of realisations of the stochastic dynamics, choosing a realistic inflationary potential tailored to create perturbations during USR that lead to PBHs with mass $7\times10^{-15} M_\odot$, roughly the mass of the asteroid Eros. After our results appeared, the form of $P(N)$ in stochastic USR, including an exponential tail, was calculated analytically \cite{Pattison:2021oen}. Our calculation in \paperI ~for the first time solved the stochastic noise self-consistently for each realisation, incorporating non-Markovian backreaction. We considered the non-linearity of the system by taking into account that the noise depends on the inflaton perturbations, while the perturbations evolve on the coarse-grained background affected by the noise. In \cite{Levasseur:2013ffa, Levasseur:2013tja, Levasseur:2014ska}, such backreaction effects had been considered recursively. However, to our knowledge, to date no calculation apart from \paperI ~has included the non-Markovian effects at every timestep, realisation by realisation. We obtained the full probability distribution, all the way from the Gaussian peak to the exponential tail for large $\Delta N$ values where PBHs form (as well as negative values of $\Delta N$). We found that in the model studied (referred to in the present paper as the asteroid case), the exponential tail led to an enhancement of the observed PBH abundance by a factor of $\sim 10^5$ over the Gaussian approximation based on the power spectrum. We also found that the non-trivial evolution of the perturbations enhances the PBH abundance by orders of magnitude, when compared to standard stochastic calculations where the mode functions are fixed to their SR super-Hubble limit, which leads to noise proportional to the Hubble parameter. Our numerical results showed that such simplifying assumptions, used in studies of stochastic inflation from the original work~\cite{Starobinsky:1986} to the most recent papers~\cite{Pattison:2021oen}, may be inaccurate in USR. We now extend this conclusion to other USR potentials.

In the present paper we return to asteroid mass PBHs, and extend our numerical treatment to other observationally interesting mass scales for PBH dark matter~\cite{Carr:2020xqk, Green:2020jor, Carr:2020gox}. We consider PBHs with a mass of $5 M_\odot$ and Planck scale relic PBHs with an initial mass of $10^3$ kg. We also consider PBHs that act as seeds for supermassive black holes in the early universe, with an initial mass of $2000 M_\odot$. As in \paperI, we build the inflationary potentials starting from the scenario where the Standard Model Higgs is the inflaton~\cite{Bezrukov:2007ep, Rubio:2018ogq}, and use the chiral Standard Model renormalisation group running to create a shallow minimum that leads to USR~\cite{Rasanen:2018fom} (see \cite{Ezquiaga:2017fvi,Ballesteros:2017fsr,Kannike:2017bxn, Bezrukov:2017dyv, Hertzberg:2017dkh} for similar constructions.) In the Higgs inflation model, a potential that produces PBHs with the right abundance (according to the Gaussian approximation) gives a spectral index that is too red on the CMB scales, except in the case of Planck mass relics. Hence, as in \paperI, we adjust the large field part of the potentials by hand to fit the CMB. Our models are therefore only inspired by Higgs inflation, not real examples of Higgs inflation (except in the relic case).

In \sec{sec:form} we go over the formalism of stochastic inflation. We highlight open issues and ambiguities in the application of the stochastic equations to the USR regime. In \sec{sec:res} we first verify our algorithm by applying it to toy models discussed in~\cite{Ezquiaga:2019ftu}. We apply our numerical machinery to more realistic potentials that simultaneously fit the CMB constraints on the observable part of the spectrum and produce PBHs with the right abundance (in the Gaussian approximation). In \sec{sec:conc} we summarise our findings. In appendix~\ref{sec:pot} we give the details of how we construct our inflationary potentials. In appendix~\ref{sec:num} we explain how we discretise and solve numerically the relevant equations. Finally, in appendix~\ref{sec:gauge} we discuss gauge issues. 

\section{Stochastic $\Delta N$ formalism} \label{sec:form}

\subsection{Coarse-graining and noise} \label{sec:Langevin}

\subsubsection{Background equations}

We consider stochastic effects in single field inflation. We follow the notation of \cite{Pattison:2019hef}. The action is
\bea \label{action}
  S = \int \rmd^4 x \sqrt{-g} \left[ \ha \Mpl^2 R - \ha g^{\a\b} \pat_\a \phi  \pat_\b \phi - V(\phi) \right] \ ,
\eea
where $\Mpl$ is the reduced Planck mass, $R$ is the Ricci scalar and $\phi$ is the inflaton. We choose units such that $\Mpl=1$. We consider a spatially flat Friedmann--Robertson--Lema\^itre--Walker universe with scalar perturbations, so the metric is
\bea
  \rmd s^2 = - ( 1 + 2 A ) \rmd t^2 + 2 a(t) \pat_i B \rmd x^i \rmd t + a(t)^2 [ ( 1 - 2 \psi ) \d_{ij} + 2 \pat_i \pat_j E ] \rmd x^i \rmd x^j \ .
\eea
We decompose the inflaton into long and short wavelength parts as
\bea \label{dec}
  \phi(t,\vec{x}) &=& \int \frac{\rmd^3 k}{(2\pi)^\frac{3}{2}} \phi_\vec{k}(t) e^{-i\vec{k} \cdot \vec{x}} \el
  &=& \int \frac{\rmd^3 k}{(2\pi)^\frac{3}{2}} \left[ 1 - W\left(\frac{k}{\sigma a H}\right) \right] \phi_\vec{k}(t) e^{-i\vec{k} \cdot \vec{x}} + \int \frac{\rmd^3 k}{(2\pi)^\frac{3}{2}} W\left(\frac{k}{\sigma a H}\right) \phi_\vec{k}(t) e^{-i\vec{k} \cdot \vec{x}} \el
  &\equiv& \phil(t,\vec{x}) + \delta\phi(t,\vec{x}) \ ,
\eea
where $k\equiv|\vec{k}|$, $H\equiv\adot/a$ is the Hubble rate (dot denotes derivative with respect to $t$), and $W\left(\frac{k}{\sigma a H}\right)$ is a window function: $W(x)\simeq0$ for $x\ll1$ and $W(x)\simeq1$ for $x\gg1$. The parameter $\sigma$ controls the separation scale $\kc \equiv \sigma aH$, and modes with $k \ll \kc$ are considered long wavelength, and those with $k \gg \kc$ are considered short wavelength. In the following, we consider patches of size $\sim \kc^{-1}$. Within each patch, the value of the long-wavelength field $\phil$ is approximately constant.

As the universe expands, modes are stretched to super-Hubble scales. As a result, the long-wavelength background over which short-scale modes evolve changes in time. Coarse-graining the short wavelength environment to get an effective theory for the long wavelength system leads to quantum noise that can be treated as classical stochastic noise if the probability distribution of the perturbations is classical. This requires the wavelength of the modes to be much larger than the Hubble radius, $\sigma \ll 1$ (we discuss classicalisation in \sec{sec:class}). The value of $\sigma$ gives an offset between the time a mode exits the Hubble radius and the time it is coarse-grained and gives a stochastic kick to the background. In SR, modes freeze in the super-Hubble regime, so the stochastic results are insensitive to the choice of $\sigma$, as long as it is sufficiently small that modes have stopped evolving~\cite{Starobinsky:1986} (but not too small \cite{Starobinsky:1994, Levasseur:2013tja, Grain:2017dqa, Tokuda:2017fdh, Tokuda:2018eqs, Pinol:2020cdp}), as the decaying mode dies out and the growing mode stops evolving quickly after the Hubble exit.

This is not the case in USR. Instead, the decaying mode turns into the growing mode, leading to a growth of the comoving curvature perturbation $\mathcal{R}$ on super-Hubble scales, and consequently to the generation of large perturbations that can potentially seed PBHs. Growth of $\mathcal{R}$ on super-Hubble scales due to violation of SR is well-known \cite{Seto:1999jc, Leach:2000yw, Leach:2001zf, Inoue:2001zt, Saito:2008em, Takamizu:2010je}, see \eg \cite{Rasanen:2018fom} for discussion of this in USR. The longer USR lasts, the larger are the scales to which the growth of $\mathcal{R}$ extends. This makes the results potentially sensitive to the choice of window function and the value of $\sigma$. Dependence of the noise on the window function has been discussed in \cite{Morikawa:1989xz}, see also \cite{Casini:1998wr, Matarrese:2003ye, Liguori:2004fa, Levasseur:2014ska, Grain:2017dqa}. The validity of the choice of $\sigma$ (more generally, the form of the Langevin equation) would ultimately have to be checked with a first principle derivation of the separation between system and environment in quantum field theory, as the results should be independent of the split, provided it is consistently applied. As it is customary, we choose a step function in momentum space, $W\left(\frac{k}{\sigma a H}\right)=\theta(k-\sigma a H)$. This corresponds to a position space window function that takes negative values, and underestimates the rate at which the two-point function falls as a function of spatial distance \cite{Winitzki:1999}. For our purposes this should not make a difference, as we do not consider spatial correlations.

We follow the usual heuristic derivation of the stochastic equations. We start from the FLRW background equations of motion for the action \re{action} \cite{Salopek:1992qy, Wands:2000dp},
\bea \label{FLRW_eom}
	\frac{\partial \phil}{\partial t} &=& \pil_t \\
	\frac{\partial \pil_t}{\partial t} &=& - 3H\pil_t - V_{,\phil} \\
	3 H^2 &=& \frac{1}{2} \pil_t^2 + V \ ,
\eea
where the subscript $t$ indicates that the field momentum is defined with respect to the time derivative. We rewrite the $t$ derivatives as derivatives with respect to the number of e-folds $N\equiv\ln a/a_*$, where $*$ refers to the time when the CMB pivot scale $k_*=0.05$ Mpc$^{-1}$ crosses the Hubble radius (note that in our convention $N$ grows with time). We then replace the FLRW field $\phil$ with the full field $\phi=\phil+\d\phi$ (where $\phil$ is not the FLRW field but the coarse-grained field). We do the same for the momentum, and denote the momentum defined with respect to the e-fold derivative by $\pi\equiv\pi_t/H$. We keep the short wavelength part in the time derivatives only, reinterpreted as noise. All in all, we get the equations
\bea \label{Langevin}
	\phil' &=& \pil + \xi_\phi \\
	\label{Langevin2}
	\pil' &=& - \left( 3 + \frac{H'}{H} \right) \pil - \frac{V_{,\phil}}{H^2} + \xi_\pi \\
	\label{eq:background_H_eom} 2 V &=& ( 6 - \bar\pi^2 ) H^2 \ ,
\eea
where prime denotes derivative with respect to $N$, and $\xi_\phi$ and $\xi_\pi$ are noise terms. It is important to use $N$ as the time coordinate when defining the noise \cite{Finelli:2008zg, Finelli:2010, Vennin:2015hra, Pattison:2019hef}. Physics is invariant under a change of coordinates, but the noise terms are not. In this treatment, they are
\bea \label{noise}
	\xi_\phi &\equiv& \int \frac{d^3 k}{(2\pi)^\frac{3}{2}} \frac{\rmd W}{\rmd N} \phi_\vec{k}(N) e^{-i\vec{k} \cdot \vec{x}} = - \int \frac{d^3 k}{(2\pi)^\frac{3}{2}} \frac{\rmd (\sigma aH)}{\rmd N}\delta(k - \sigma a H) \d\phi_\vec{k}(N) e^{-i\vec{k} \cdot \vec{x}} \\%\el
	\label{noise2}
	\xi_\pi &\equiv& \int \frac{d^3 k}{(2\pi)^\frac{3}{2}} \frac{\rmd W}{\rmd N} \phi'_\vec{k}(N) e^{-i\vec{k} \cdot \vec{x}} = - \int \frac{d^3 k}{(2\pi)^\frac{3}{2}} \frac{\rmd (\sigma aH)}{\rmd N}\delta(k - \sigma a H) \d\phi'_\vec{k}(N) e^{-i\vec{k} \cdot \vec{x}} \, .
\eea
The average over the noise is zero, and its two-point function is obtained by quantising $\phi$. We write
\bea \label{phioperator}
	\hat{\phi}_\vec{k} &\equiv& \phi_k \hat{a}_\vec{k} + \phi^*_k \hat{a}^\dagger_\vec{-k} \\%\el
	\label{phioperator2}
	\hat{\pi}_\vec{k} &\equiv& \phi'_k \hat{a}_\vec{k} + \phi'^*_k \hat{a}^\dagger_\vec{-k} \ ,
\eea
where the creation and annihilation operators obey the usual commutation relations
\begin{equation} \label{ladder}
	\qty[\hat{a}_\vec{k}, \hat{a}^\dagger_\vec{k'}] = \delta^{(3)}(\vec{k} - \vec{k'}) \ , \quad \qty[\hat{a}_\vec{k}, \hat{a}_\vec{k'}] = \qty[\hat{a}^\dagger_\vec{k}, \hat{a}^\dagger_\vec{k'}] = 0 \ ,
\end{equation}
and the vacuum state satisfies $\hat{a}_\vec{k}\ket{0} = 0$.

The resulting quantum noise correlators from \re{noise}--\re{ladder} are:
\bea  \label{noise_correlator}
  \expval{\hat{\xi}_\phi(N) \hat{\xi}_\phi(N')}{0} &=& \frac{1}{6\pi^2} \frac{\rmd (\sigma aH)^3}{\rmd N} |\d\phi_{k=\sigma aH}|^2 \delta(N-N') \\
  \label{noise_correlator2}
  \expval{\hat{\xi}_\pi(N) \hat{\xi}_\pi(N')}{0} &=& \frac{1}{6\pi^2} \frac{\rmd (\sigma aH)^3}{\rmd N} |\d\phi'_{k=\sigma aH}|^2 \delta(N-N') \\
  \expval{\hat{\xi}_\phi(N) \hat{\xi}_\pi(N')}{0} &=&
  \label{noise_correlator3}
  \frac{1}{6\pi^2} \frac{\rmd (\sigma aH)^3}{\rmd N} \d\phi_{k=\sigma aH} \d\phi'^*_{k=\sigma aH} \delta(N-N') \ .
\eea
The physical reason for the noise is that, unlike in standard perturbation theory, the split between the background and perturbations is time-dependent, and modes cross from small wavelengths into large wavelengths. (As $\sigma\ll1$, rather than small and large wavelength modes, it would be more accurate to talk about infrared and far infrared modes.) Choosing the window function to be a step function offers considerable simplification, as each mode contributes to the noise only when it crosses the coarse-graining radius $1/(\sigma a H)$, and \re{Langevin}, \re{Langevin2} become Langevin equations. For other window functions the noise involves correlations between different $k$ values.

There are several detailed analyses of the stochastic equations, including derivations from and comparisons to quantum field theory \cite{Morikawa:1989xz, Mijic:1994vv, Habib:1992ci, Tsamis:2005hd, Woodard:2005cv, vanderMeulen:2007ah, Finelli:2008zg, Beneke:2012kn, Levasseur:2013ffa, Gautier:2013aoa, Garbrecht:2013, Garbrecht:2014dca, Levasseur:2014ska, Burgess:2014eoa, Burgess:2015ajz, Onemli:2015, Boyanovsky:2015tba, Boyanovsky:2015jen, Vennin:2015hra, Moss:2016uix, Collins:2017haz, Prokopec:2017vxx, Tokuda:2017fdh, Tokuda:2018eqs, Pinol:2020cdp, Pattison:2019hef}. However, none with explicit Langevin equations apply to the USR case, where the results can be sensitive to the choice of the window function, which for the step function reduces to the choice of $\sigma$. The value of $\sigma$ has to be sufficiently small for the probability distribution of the short modes to have become classical. Beyond that, different choices correspond to different splits between the environment and the system, and hence to different stochastic equations. It is likely that the correct value for the equations \re{Langevin}--\re{Langevin2} and \re{noise_correlator}--\re{noise_correlator3} cannot be determined heuristically, but requires a quantum field theory calculation applied to the USR case. A proper derivation could also modify the form of the Langevin equation beyond the choice of $\sigma$. Such analysis goes beyond the scope of our paper, and we simply take a step function and choose a value of $\sigma$ for which the probability distribution of the modes has become sufficiently classical. We also consider smaller and larger values of $\sigma$, and find that the results are rather insensitive to the precise value.

\subsubsection{Perturbation equation}

We also have to consider the evolution of the modes, which determine the amplitudes of the stochastic kicks. In most explicit studies of stochastic inflation, the modes are taken to evolve on a non-stochastic FLRW background. Furthermore, the mode functions are fixed to their asymptotic SR and super-Hubble limit, $|\d\phi_{\vec k}|=H/(\sqrt{2}k^{3/2})$, and the time derivative of $H$ in the noise amplitude \re{noise_correlator} is neglected. The momentum noise then vanishes, and the field noise correlator becomes
\bea \label{noise_correlator_simplified}
  \expval{\hat{\xi}_\phi(N) \hat{\xi}_\phi(N')}{0} &=& \left(\frac{H}{2\pi}\right)^2 \delta(N-N') \ ,
\eea
where $H$ is the FLRW Hubble rate. This simplifies the problem considerably, because there is no need to follow the evolution of the modes, and the stochastic process becomes Markovian. However, for consistency, the backreaction of non-trivial mode evolution and change of the environment should be included, particularly in USR where modes can evolve after Hubble crossing. This has been done recursively in \cite{Levasseur:2013ffa, Levasseur:2013tja, Levasseur:2014ska}, and consistently at every timestep in \paperI.

In standard linear perturbation theory, the short wavelength modes obey linear perturbation equations over the FLRW background \cite{Salopek:1992qy, Wands:2000dp}. In the spatially flat gauge ($\psi=E=0$), these equations reduce to\footnote{Previous versions of the paper had a typo in the last term, with $\pil$ instead of $\pil_t$. This does not affect the numerical results.}
\begin{equation} \label{eq:field_mode_eom}
	\frac{\pat^2\d\phi_\vec{k}}{\pat t^2} + 3H \frac{\pat \d\phi_\vec{k}}{\pat t} + \qty[ \frac{k^2}{a^2} + V''(\phil) - \frac{1}{a^3} \frac{\rmd}{\rmd t}\qty(\frac{a^3}{H}\pil_t^2) ]{\d\phi}_\vec{k} = 0 \ ,
\end{equation}
where the barred symbols are FLRW background quantities, and $\d\phi_\vec{k}$ is the perturbation around the FLRW field (or the mode function in the quantum operator \eqref{phioperator}). We include the change of the coarse-grained background in the evolution of the small wavelength modes by assuming that the perturbation $\d\phi_\vec{k}$ follows \re{eq:field_mode_eom} with the FLRW fields replaced by the coarse-grained fields. Switching to derivatives with respect to the number of e-folds, the resulting equation for the short wavelength modes is\footnote{Previous versions of the paper had a typo in the $\pil^2$ term of $\omega_k^2$. This does not affect the numerical results.}
\begin{eqnarray} \label{eq:field_mode_eom_consistent}
&&~~~~~~~~~~~~~~~~~~~~\delta\phi_{\vec{k}}'' + \left(3  + \frac{H'}{ H}\right) \delta\phi_{\vec{k}}' + \omega_k^2 \delta\phi_{\vec{k}} = 0 \ ,~~~~~~~~~~~~~~\\
\hspace*{-1.0cm}{\rm where} &&\nonumber\vspace*{-0.5cm}\\
&&~~~~~~~~~~~~~~~~~~~~\omega_k^2 \equiv {k^2\over(a H)^2} +\pil^2 \left(3 + {H'\over H}\right) + 2\pil {V_{,\phil}\over H^2} + {V_{,\phil\phil}\over H^2}\,.~~~~~~~~~~~~~~    
\end{eqnarray}

As in standard perturbation theory, we assume that the initial conditions in the deep sub-Hubble regime are given by the Bunch--Davies vacuum (we discuss the vacuum in more detail in \sec{sec:class}),
\begin{equation} \label{eq:bunch_davies}
	\d\phi_\vec{k} = \frac{1}{\sqrt{2k} a} \ , \qquad (a \d\phi_\vec{k})' = - i \frac{k}{H} \d\phi_\vec{k} \ .
\end{equation}

\subsubsection{Remarks} \label{sec:remarks}

A few comments are in order. In \re{eq:background_H_eom} we have chosen the first Friedmann equation not to receive noise corrections. This is usual in stochastic studies. However, as the continuity equation (\ie the field equation of motion) has noise, both Friedmann equations cannot be noise-free. Therefore, given our choice for the first Friedmann equation, the second Friedmann equation must have noise. One reason for our choice is that the first Friedmann equation written in terms of $H$ does not involve time derivatives, which in the heuristic derivation of the Langevin equation are the origin of the noise. Still, in order to derive the second order field equation \re{eq:field_mode_eom} from the perturbed Friedmann equations, it is necessary to use the second Friedmann equation. If we added noise directly to the Einstein equations perturbed around the FLRW universe, and derived the second order field equation by taking derivatives as usual, the resulting stochastic counterpart to the mode equation \re{eq:field_mode_eom_consistent} would involve noise explicitly. Correspondingly, the noise amplitudes \re{noise_correlator} and \re{noise_correlator2} would themselves depend on the noise. We choose to avoid such complications, and take factors of $H'$ appearing in the perturbation equation to be related to the coarse-grained fields, in the same way as in the FLRW case, $H'/H=-\frac{1}{2}\pil^2$. While the noise correction to $H'/H$ can be one order of magnitude larger than the classical contribution (which is strongly suppressed in USR), it is always orders of magnitude smaller than unity. Because $H'/H$ only appears in the combination $3+H'/H$, we expect this noise correction to have negligible impact on the calculation.

In addition to the full non-Markovian calculation, for the sake of comparison we also do a simplified calculation where the kicks are given by \re{noise_correlator_simplified}, corresponding to the asymptotic SR and long-wavelength behaviour of the mode functions (and also neglecting a time derivative of $H$). We find that this approximation, commonly used in the literature, severely underestimates or overestimates the PBH abundance, depending on the potential. We discuss the details in \sec{sec:real}. We also use a recursive treatment where the modes are first taken to evolve on the FLRW background, and the noise given by these modes determines the coarse-grained background for the next iteration, as in \cite{Levasseur:2013ffa, Levasseur:2013tja, Levasseur:2014ska}. We find that one iteration is sufficient to reproduce the results of the full treatment.

As the stochastic formalism goes beyond perturbation theory, the correct equations cannot be reliably deduced using the heuristic arguments we have given, but should be derived from quantum field theory. Our prescription amounts to using the equations for perturbations around homogeneous and isotropic background also for perturbations around the coarse-grained background field. There could be couplings between the perturbations and the modes that form the mean field that our treatment does not capture. Note that while all the modes counted as part of the background are super-Hubble and classical, the background is not exactly homogeneous and isotropic. Modes with $k$ close to the Hubble scale will significantly evolve again during USR, while modes with smaller $k$ will not, leading to a time-dependent gradient in the background. 

We also have to consider gauge-dependence. The stochastic equations \re{Langevin}--\re{Langevin2} for $\phil$ have been formulated in the uniform $N$ gauge, whereas the equation of motion \re{eq:field_mode_eom_consistent} for the perturbations is in the spatially flat gauge. It was shown in \cite{Pattison:2019hef} that the correction from the gauge transformation is small both in SR and deep in USR. However, this is not guaranteed to be the case during the transition between the two regimes, or when the potential makes a sizeable contribution during USR. We have verified numerically that for the inflationary potentials we use to produce asteroid, solar mass and supermassive PBHs, the correction from the gauge transformation, computed as the ratio of the correction over the mode amplitude, is less than $10^{-5}$ for most modes. There is, however, a narrow spike of modes with very small amplitude at the coarse-graining scale, for which the gauge correction rises to a maximum of $\sim10^{-2}$. These modes exit the coarse-graining scale roughly halfway in the USR region, when the field has just passed the local minimum in the potential. The contribution of these modes to the overall dynamics is small. For details, see appendix~\ref{sec:gauge}. In the relic case, the situation is different and the gauge transformations have a large impact also on modes outside the spikes. As that case also suffers from other technical and conceptual problems, we postpone discussion of this issue to \sec{sec:relic}. 

Note that the uniform $N$ gauge has a residual gauge freedom, which is not reflected in the Langevin equation. Under such a residual gauge transformation, the perturbation equations change, while the Langevin equation does not, which is in principle inconsistent. Furthermore, we have only considered linear order gauge transformations, whereas the stochastic calculation contains non-perturbative physics. The system of equations of large and small wavelengths as a whole should be gauge-invariant, and a derivation from first principles in quantum field theory should determine its correct form. Such a derivation goes beyond the scope of this paper.

Finally, defining stochastic equations involves specifying the discretisation, for example choosing between the It\^o and the Stratonovich prescription. This issue for stochastic inflation is discussed in detail in \cite{Pinol:2018, Pinol:2020cdp}, but it is unimportant for single field inflation with a sub-Planckian potential, $V\ll1$, such as ours. Our exact numerical implementation, including the discretisation details of the evolution equations, are given in appendix \ref{sec:num}.

\subsection{Squeezing, freezing, and classicalisation} \label{sec:class}

As noted earlier, in order for the stochastic formalism to apply, the coarse-graining parameter $\sigma$ has to be small enough that the probability distribution of the perturbations has become classical by the time they join the background. As modes cross the Hubble scale, they are squeezed, and the expectation values of powers of fields and momenta approach those of a classical stochastic distribution \cite{Guth:1985ya, Grishchuk:1989ss, Grishchuk:1990bj, Prokopec:1992ia, Albrecht:1992kf, Martin:2007bw, Martin:2012ua, Grain:2019vnq}. Squeezing is due to the rapid decline of the decaying mode of the perturbation, and leads to what has been called ``decoherence without decoherence'' \cite{Polarski:1995jg}, since it leads to classicalisation in the sense explained above, but not to proper decoherence, \ie suppression of the off-diagonal elements of the density matrix \cite{Guth:1985ya, Giulini:1996nw, Kiefer:1998qe, Kiefer:1998jk, Martineau:2006ki, Burgess:2006jn, Kiefer:2006je, Kiefer:2007zza, Kiefer:2008ku, Ye:2018kty}. Therefore, squeezing captures only part of decoherence and classicalisation, and the entire process is not fully understood \cite{Perez:2005gh, Sudarsky:2009za, Martin:2012pea, Crull:2014ifa, Okon:2015fgr, Ashtekar:2020mdv, Berjon:2021umq}. We will be content to use the degree of squeezing as a criterion for classicalisation, as in \cite{Polarski:1995jg}. This approach can be considered conservative in the sense that any decoherence due to interactions is expected to only increase the degree of classicality.

To simplify the discussion and make contact with the standard squeezing formalism, in this section we work with a discrete set of $\vec{k}$ modes instead of a continuum, replacing Fourier transforms such as \eqref{dec} with Fourier series, and Dirac delta functions such as those in the ladder commutator \eqref{ladder} with Kronecker deltas. This corresponds to working in a finite box with periodic boundary conditions. The operators $\hat\phi_{\vec{k}}$, $\hat\pi_{\vec{k}}$ mix the $\vec{k}$ and $-\vec{k}$ modes and are non-Hermitian. We obtain two independent Hermitian field operators for each $(\vec{k}, -\vec{k})$ pair by defining the real and imaginary parts $\hat\phi^{R}_{\vec{k}} \equiv \frac{1}{\sqrt{2}}(\hat\phi_{\vec{k}} + \hat\phi_{\vec{k}}^\dagger)$, $\hat\phi^{I}_{\vec{k}} \equiv -\frac{i}{\sqrt{2}}(\hat\phi_{\vec{k}} - \hat\phi_{\vec{k}}^\dagger)$, and similarly for $\hat\pi_{\vec{k}}$ \cite{Martin:2007bw}. They behave identically, and in the following we concentrate on one of them, dropping the superscript for brevity.

In the Schr\"odinger picture, a squeezed state can be written as (we omit a phase which does not contribute to the expectation values we consider) \cite{Polarski:1995jg,Grain:2019vnq}
\begin{eqnarray} \label{psi}
    \ket{\psi_\vec{k}} = \exp[\frac{1}{2}\qty(s_\vec{k}^*\hat{c}_\vec{k}^2 - s_\vec{k}\hat{c}_\vec{k}^\dagger{}^2)] \ket{0} \ ,
\end{eqnarray}
where $s_\vec{k}=r_k e^{2i\varphi_\vec{k}}$ is the squeezing parameter and $\hat{c}^\dagger_\vec{k}$ and $\hat{c}_\vec{k}$ are creation and annihilation operators. The amplitude $r_k$ indicates how squeezed the state is, and the phase $\varphi_\vec{k}$ gives the squeezing direction in phase space. The $\hat{c}_\vec{k}$ operators determine the vacuum state with respect to which squeezing is measured. They are related in the standard way to the system's canonical variables $\hat{q}_{\vec{k}}$ and $\hat{p}_{\vec k}$, which we choose to be
\bea \label{eq:QP_decomp}
    \hat{q}_{\vec{k}} &=& \frac{1}{\sqrt{2}}\qty(\hat{c}_{\vec{k}} + \hat{c}^\dagger_{\vec{k}}) = \sqrt{k}a \hat\phi_{\vec{k}} \\ \label{eq:QP_decompII}
   \hat{p}_{\vec k} &=& -\frac{i}{\sqrt{2}}\qty(\hat{c}_{\vec{k}} - \hat{c}^\dagger_{\vec{k}}) = \frac{H a^2}{\sqrt{k}} \hat\pi_{\vec{k}}\ .
\eea
Deep inside the Hubble radius the state annihilated by $\hat{c}_{\vec{k}}$ is the adiabatic Bunch--Davies vacuum, and the creation and annihilation operators coincide with $\hat{a}^\dagger_\vec{k}$ and $\hat{a}_\vec{k}$ in \eqref{phioperator} (up to the normalisation difference between the continuum and discrete case). When $k \lesssim aH$, this is no longer the case; in fact, there is no adiabatic vacuum at all. The choice \re{eq:QP_decomp}--\re{eq:QP_decompII} gives the closest match to the usual discussion of squeezing for the Sasaki--Mukhanov variable, which in the spatially flat gauge reduces to $a \phi_{\vec{k}}$.\footnote{The Bunch--Davies vacuum does not uniquely fix the canonical variables. If we add terms that are subleading in $(a H)/k$, they vanish in the deep sub-Hubble limit where the vacuum is defined, but can be important for super-Hubble wavelengths. Our choice is consistent with the Bunch--Davies vacuum, but not determined by it.}

From \re{psi} and \re{eq:QP_decomp}--\re{eq:QP_decompII} we get the correlators \cite{Grain:2019vnq}
\begin{eqnarray} \label{eq:r_in_QP}
  \Sigma_{_{\hspace*{-0.4mm}qq}}(\vec{k}) &\equiv& \expval{\hat{q}_{\vec{k}}^2}{\psi_{\vec{k}}} = \frac{1}{2}\qty[\cosh(2 r_k) - \cos (2\varphi_k) \sinh(2 r_k)] = ka^2|\phi_k|^2 \\%\el
  %\sigma_{PP\,\vec{k}}
  \label{eq:r_in_QPII}
  \Sigma_{_{\hspace*{-0.4mm}pp}}(\vec{k}) &\equiv& \expval{\hat{p}_{\vec{k}}^2}{\psi_{\vec{k}}} = \frac{1}{2}\qty[\cosh(2 r_k) + \cos(2\varphi_k) \sinh(2 r_k)] = \frac{H^2a^4}{k}|\phi'_k|^2 \\%\el
  %\sigma_{QP\,\vec{k}} 
  \label{eq:r_in_QPIII}
  \Sigma_{_{\hspace*{-0.4mm}qp}}(\vec{k}) &\equiv& \frac{1}{2}\expval{ \hat{q}_{\vec{k}} \hat{p}_{\vec{k}} + \hat{p}_{\vec{k}} \hat{q}_{\vec{k}}}{\psi_{\vec{k}}} = - \frac{1}{2}\sin(2\varphi_k) \sinh(2 r_k) = a^3 H\Re(\phi_k^*\phi'_k) \ ,
\end{eqnarray}
where we have made contact with the Heisenberg picture formalism by expressing the correlators in terms of the mode functions $\phi_k$ used in \eqref{phioperator}. These expectation values are generically large if the squeezing amplitude $r_k$ is large. However, for particular values of the squeezing angle $\varphi_k$, some of the expectation values are small even for large $r_k$. The sum $\Sigma_{_{\hspace*{-0.4mm}qq}}(\vec{k}) + \Sigma_{_{\hspace*{-0.4mm}pp}}(\vec{k}) = \cosh(2 r_k)$  is independent of $\varphi_k$, so it is always large when the state is very squeezed. 

For the Bunch--Davies vacuum, the mode is initially described by the minimum uncertainty wave packet, for which $r_k=0$. As the mode stretches further outside the Hubble radius, the phase space probability distribution gets squeezed and $r_k$ grows. Because $n_{\vec k} \equiv \expval{\hat{c}^\dagger_{\vec{k}} \hat c_\vec{k}}{\psi_{\vec{k}}}=\ha [1+\cosh(2 r_k)]$, the condition $r_k\gg1$ corresponds to large occupation number $n_{\vec k} \gg 1$. For large $r_k$, the probability distribution covers a large region in phase space where the expectation values of the even powers of the fields are large compared to the expectation value of the commutator $[\hat{q}_{\vec{k}},\hat{p}_{\vec{k}}] = i$. Therefore, all relevant expectation values can be well approximated by a classical probability distribution.

Importantly for the stochastic calculation, squeezing also makes the operators $\hat{q}_{\vec{k}}$ and $\hat{p}_{\vec{k}}$ highly correlated, so the field and momentum kicks become proportional to each other. This can be seen from the classical Gaussian probability distribution that reproduces the correlators \eqref{eq:r_in_QP}--\eqref{eq:r_in_QPIII}, together with $\langle\hat{q}_{\vec{k}}\rangle = \langle\hat{p}_{\vec{k}}\rangle = 0$. (This approach can be compared to analyses that use the Wigner function \cite{Kiefer:1998jk, Martin:2007bw, Kiefer:2008ku, Martin:2012ua, Grain:2019vnq}.) Dropping the subscript $\vec{k}$ to lighten the notation, the distribution is
\begin{equation}
\label{eq:gaussian_probability_distribution}
	P(q, p) = \frac{1}{2\pi\sqrt{\Sigma_{qq}\Sigma_{pp} - \Sigma_{qp}^2}}
	\exp[-\frac{ \Sigma_{pp}q^2 + \Sigma_{qq}p^2 - 2\Sigma_{qp}qp }{2(\Sigma_{qq}\Sigma_{pp}-\Sigma_{qp}^2)} ] \ .
\end{equation}
From \eqref{eq:r_in_QP}--\eqref{eq:r_in_QPIII}, the squeezing limit $r_k \to \infty$ corresponds to $\Sigma^2_{qp} \to \Sigma_{qq}\Sigma_{pp}$. In this limit, the distribution becomes
\begin{equation} \label{eq:squeezed_p}
	P(q,p) \to \frac{1}{\sqrt{2\pi\Sigma_{qq}}} e^{-\frac{q^2}{2\Sigma_{qq}}} \delta\qty(p - \sign(\Sigma_{qp})\sqrt{\frac{\Sigma_{pp}}{\Sigma_{qq}}}q) \ .
\end{equation}
The variables $q$ and $p$ become tightly correlated, and the phase space distribution is squeezed to the line $p = \sign(\Sigma_{qp})\sqrt{\frac{\Sigma_{pp}}{\Sigma_{qq}}}q$. This also applies to $\phi_\vec{k}$ and $\phi'_\vec{k}$ and the kicks, giving $\xi_\pi = \xi_\phi \Re \frac{\phi'_k}{\phi_k}$.\footnote{To see this, first note that in the squeezed limit $\Sigma^2_{qp} \to \Sigma_{qq}\Sigma_{pp}$, \eqref{eq:r_in_QP}--\eqref{eq:r_in_QPIII} give $|\Re (\phi_k^* \phi_k')| = |\phi_k^* \phi_k'|$. This is only possible if $\Re (\phi_k^* \phi_k') = \phi_k^* \phi_k'$, that is, $\phi_k^* \phi_k'$ is real. Then
\begin{equation*}
    \sign(\Sigma_{qp})\sqrt{\frac{\Sigma_{pp}}{\Sigma_{qq}}}
    = \frac{\phi_k^* \phi_k'}{|\phi_k^* \phi_k'|}\frac{|\phi'_k|}{|\phi_k|}
    = \frac{\phi'_k}{\phi_k} = \Re \frac{\phi'_k}{\phi_k} \ .
\end{equation*}
The last equality follows from the reality of $\phi_k^* \phi_k'$ by noticing that $\phi'_k / \phi_k = |\phi'_k|^2/(\phi_k^* \phi_k')^* \in \mathbb{R}$. Since the state is not completely squeezed in our practical computations, we take the real part to get rid of the small imaginary contribution.} Therefore we effectively have only one random variable. In SR, this variable is aligned almost purely with the $\phi$ direction, as discussed around \eqref{noise_correlator_simplified}. Outside of SR, the $\pi$ component can, in principle, become significant, though it remains correlated with the $\phi$ component.

We use the value of $r_k$ as a measure of classicalisation, and demand that modes satisfy $\cosh(2r_k)>100$ before they are counted as part of the background. Setting this as both the necessary and sufficient condition would lead to a time-dependent coarse-graining parameter $\sigma$. We skip such a complication, as our treatment of classicalisation is in any case simplified. Instead, we choose a constant value of $\sigma$ small enough that the condition $\cosh(2r_k)>100$ is satisfied for all modes before they are coarse-grained. With the exception of the Planck mass relic dark matter case, the choice $\sigma=0.01$ is sufficient to guarantee this. Slightly larger values of $\sigma$ could also be used, up to $\sigma = 0.07$ for the asteroid case, and $\sigma = 0.03$ for the solar and the supermassive case. We also consider smaller values and find that the results remain pretty much the same, unless very small values of $\sigma$ are considered, so that the amplitude of the kicks changes significantly between the Hubble exit and coarse-graining. For the Planck mass relic case we have to take $\sigma=0.35$ for the modes that produce PBHs to reach the classicalisation threshold. We discuss the dependence on the value of $\sigma$, and the problems of the relic case, in \sec{sec:res}.

\begin{figure}[t!]
\centering
\includegraphics{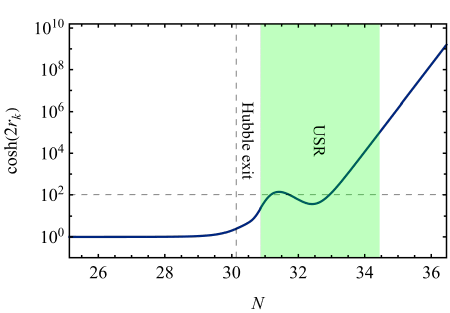}
\includegraphics{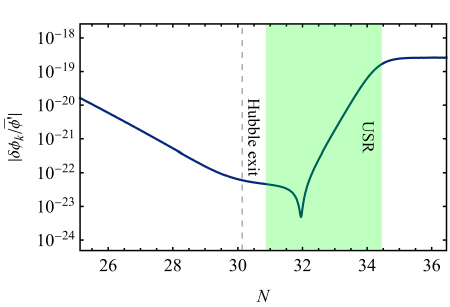}
\caption{Squeezing parameter (left) and perturbation amplitude $\delta\phi_k/\phil'$ (right) as a function of e-folds $N$ in the asteroid case for a mode that exits the Hubble radius shortly before USR. The dashed vertical line marks the Hubble exit, and the green area indicates the USR region.}
\label{fig:r}
\end{figure}

A possible complication arises from the fact that $r_k$ is not always a monotonic function of time. This is not a problem in SR, where modes only get more squeezed with time. However, when SR is broken, squeezing can be reduced, even completely undone. In \fig{fig:r} (left) we show $r_k$ for a single mode as a function of time for the asteroid case. The mode exits the Hubble radius at $N=30.1$, and inflation ends at $N=51.7$. The field is in USR from $N=30.9$ to $N=34.5$. Squeezing grows until a few e-folds after Hubble exit, and the classicality threshold $\cosh(2r_k)=100$ is reached at $N=31.2$. However, after $N=31.4$, the squeezing amplitude $r_k$ starts to decrease, and the mode goes under the threshold at $N=31.7$, before crossing back at $N=32.9$. The situation does not change if we change the classicality threshold $r_k$, as some other modes will cross back under the new threshold. To get insight into the origin of this feature, we show in \fig{fig:r} (right) the normalised field perturbation $\delta{\phi}_k/\phil'$, which, like the power spectrum $\PR(k)$, freezes to a constant value during SR on super-Hubble scales. The modes for which squeezing is partly undone exit the Hubble radius just before or during the USR regime. This is why the plot does not have a constant part at early times: the mode transitions from the standard sub-Hubble time evolution directly to USR time evolution. The plateau on the right is the super-Hubble SR regime after USR. We see that the decrease in $r_k$ is related to (but does not perfectly correlate with) a sharp dip of $\delta{\phi}_k/\phil'$ followed by rapid growth. Such behaviour is generic in transitions between SR and USR \cite{Byrnes:2018txb, Passaglia:2018ixg, Carrilho:2019oqg, Ozsoy:2019lyy, Tasinato:2020vdk}.

We simply assume that once modes cross our classicality threshold for the first time, they irreversibly become part of the classical stochastic background. In our simulations, 1\% to 8\% of the modes cross back and are desqueezed, depending on the potential and stochastic realisation. These modes exit the Hubble radius shortly before USR, and reach the coarse-graining scale soon after the end of USR, when the background field is still near the feature in the potential. We checked that disregarding this small range of modes does not have a noticeable impact on the results. Note that in addition to squeezing, classicalisation also proceeds through decoherence via interactions, which there is no reason to expect to be undone by USR. For this reason, treating classicalisation as irreversible seems justified. Ultimately, this issue would have to be addressed by properly treating the division into classical background and quantum perturbations from first principles in quantum field theory.

\subsection{Stochastic $\Delta N$ formalism and black hole abundance} \label{sec:deltan}

Stochastic effects lead to an exponential tail \cite{Pattison:2017mbe, Ezquiaga:2019ftu, Figueroa:2020jkf} for the probability distribution $P(\R)$ of the comoving curvature perturbation $\R$. It is therefore not enough to consider the power spectrum $\mathcal{P}_{\R}(k)$ to calculate the PBH abundance. We instead use the $\Delta N$ formalism \cite{Sasaki:1995aw, Sasaki:1998ug, Wands:2000dp, Lyth:2004gb} to find the value of $\R$ on a coarse-grained patch, given by
\begin{equation}\label{eq:DeltaNdef}
    \R = N - \bar{N} \equiv \Delta N \, .
\end{equation}
Here $N$ is the number of e-folds in the local patch from an initial unperturbed hypersurface with constant initial field value $\phii$ to a final hypersurface with constant field value $\phif$ that corresponds roughly to the end of inflation, and $\bar{N}$ is the mean number of e-folds averaged over stochastic realisations (\ie patches). Choosing a final hypersurface with constant field value $\phif$ corresponds to calculating the curvature perturbation in the comoving gauge, where field perturbations vanish.

The patch size is determined by the coarse-graining scale $\kc=\sigma a H$. We start the simulation when the wavenumber of the mode corresponding to the CMB pivot scale $k_*$ equals $100 aH$. We have checked that starting the simulation earlier to accumulate kicks from a longer period has negligible impact on our results, as the kicks during the SR stage are small and far removed from the PBH scale. We tested this with 50, 100 and 500 e-folds of evolution before the CMB pivot scale. Our results do not exclude the possibility that there are effects that grow logarithmically or otherwise very slowly with $N$, in which case the history of kicks could be important if inflation lasts very long before the scale $k_*$ exits the Hubble radius. Also, if inflation lasts long, expectation values may not be representative of the observations made in our universe over a finite volume \cite{Nurmi:2013xv}.

We stop the stochastic kicks when the patch reaches the scale $\kpbh$ that corresponds to the PBH mass we consider, given below in \re{eq:M_PBH}. The reason is that perturbations with wavelengths smaller than the size of the collapsing region do not affect PBH formation at this scale: they behave as noise that averages out in the coarse-graining process. (For the asteroid case we have explicitly checked that continuing the kicks until the end of inflation and averaging over scales below the PBH scale gives the same result as simply stopping the kicks when $\kc = \kpbh$.) Subsequently, we evolve the equations without kicks until $\phif$ is reached. We record the final value of $N$, and repeat the process many times to find the probability distribution $P(N)$, and thus $P(\R)$.

This approximation does not take into account correlations between different PBHs. Two overdensities separated by distance $L$ have shared the same kicks from modes with wavelengths larger than $L$, but have a different history of kicks from shorter wavelength modes. See \cite{Ando:2020fjm} for a discussion of how to obtain the two-point function from the Langevin equations. We consider only PBH abundance, \ie the one-point function, not spatial correlations.

Sometimes the uniform density curvature perturbation $\zeta$ is used instead of $\R$. In SR, these perturbations are equal up to a sign, but during USR they evolve differently, see \eg \cite{Malik:2008}. However, if we have a period of SR after USR, it doesn't make a difference which one we use. If USR is not followed by SR, or the SR period is short (as in the case of relic PBHs), $\zeta$ and $\R$ may not have had time to equalise, and one should be careful regarding which curvature perturbation to use and how to relate it to the density contrast that determines PBH formation.

When a sufficiently overdense perturbation of wavenumber $k$ re-enters the Hubble radius during the radiation-dominated phase after inflation, almost the entire Hubble radius collapses into a PBH with mass \cite{Rasanen:2018fom, Carr:1975qj}
\begin{eqnarray} \label{eq:M_PBH}
    M = \frac{4}{3}\pi \gamma H_k^{-3} \rho_k \approx 5.6 \times 10^{15} \gamma \left(\frac{k}{k_*}\right)^{-2} M_\odot \, ,
\end{eqnarray}
where $M_\odot \approx 2 \times 10^{33}$ g, $\gamma \approx 0.2$ is a parameter that characterises the collapse \cite{Carr:1975qj}, and $H_k$ and $\rho_k$ are the background Hubble rate and the energy density at Hubble entry, respectively. We assume standard radiation-dominated expansion history with the Standard Model degrees of freedom.

There are various analyses of and refinements to the treatment of the collapse \cite{Carr:1975qj, Niemeyer:1999ak, Shibata:1999zs, Musco:2004ak, Polnarev:2006aa, Musco:2012au, Harada:2013epa, Nakama:2013ica, Young:2014ana, Harada:2015yda, Motohashi:2017kbs, Yoo:2018, Germani:2018jgr, Musco:2018rwt, Kawasaki:2019mbl, DeLuca:2019qsy, Young:2019yug, Young:2019osy, Kehagias:2019eil, Escriva:2019nsa, Escriva:2019phb, Germani:2019zez, Suyama:2019npc, Young:2020xmk, Yoo:2020lmg, Wu:2020ilx, Tokeshi:2020tjq, Escriva:2020tak, Musco:2020jjb, Yoo:2020dkz, Biagetti:2021eep, Kitajima:2021fpq, Wang:2021kbh}. We simply assume that a region collapses if the comoving curvature perturbation exceeds a critical value $\R_c$. We adopt the value $\R_c=1$, close to the most recent estimates. The abundance is extremely sensitive to this choice. Increasing or decreasing $\R_c$ by 10\% changes the abundance by $\sim$ 1--2 orders of magnitude in the asteroid, solar and supermassive cases; the relic case is more stable. Decreasing to the original estimate $\R_c=3/4$ from \cite{Carr:1975qj}\footnote{This corresponds to the critical density perturbation $\delta_c=1/3$, equal to the equation of state parameter $w$ in radiation domination, related to the comoving curvature perturbation by $\delta = -\frac{2(1+w)}{5+3w}\nabla^2\R/(aH)^2$, or $\R_c = \frac{9}{4}\delta_c$ at Hubble crossing (see \eg \cite{Liddle:2000cg}).} would increase the abundance by 3, 3, and 5 orders of magnitude in the asteroid, solar and supermassive cases, respectively, and by less than an order of magnitude in the relic case. This issue concerns the correct modelling of the PBH formation process, given the probability distribution $P(\R)$. Therefore, we do not consider it further, since our aim is to determine the shape of $P(\R)$, and see how it affects the abundance, given a procedure to calculate the abundance. Furthermore, a tiny change in the parameters of the inflationary potential leads to a large difference in the PBH abundance, which can compensate for our simplified modelling of the collapse.

The fraction of simulations where $\R > \R_c$ gives the initial fraction $\beta$ of the energy density in PBHs,
\bea \label{beta}
  \b = 2 \int_{\Delta N>\R_c}^\infty \rmd N P(N) \ ,
\eea
where the factor of 2 comes from the Press--Schechter prescription to solve the cloud-in-a-cloud problem \cite{Press:1973iz}. Since the energy density of PBHs scales as non-relativistic matter, their fraction of the energy density grows during radiation domination, and today it is~\cite{Rasanen:2018fom}
\begin{equation} \label{eq:omega_PBH}
    \Omega_\text{PBH} \approx 9 \times 10^7 \gamma^{\frac{1}{2}} \beta \left(\frac{M}{M_\odot}\right)^{-\frac{1}{2}} \ .
\end{equation}
In order not to overproduce dark matter, the initial abundance $\b$ has to be small, as the collapse happens deep in the radiation-dominated era, and hence the energy density fraction has a lot of time to increase. In the cases we consider below, $\beta$ varies between $10^{-17}$ and $10^{-8}$.

If the distribution of $\R_k$ were Gaussian \cite{Green:2004wb, Young:2014ana}, as is often assumed in the literature, its variance would be $\sigma_\R^2 = \int_{k_\mr{IR}}^{\kpbh} \rmd (\ln k) \PR(k)$, where $\PR$ is the curvature power spectrum and ${k_\mr{IR}}$ is a cutoff corresponding roughly to the present Hubble radius. (The precise value of $k_\mr{IR}$ makes no difference.) The initial energy density fraction is then
\begin{equation} \label{eq:gaussian_beta_0}
    \beta = 2\int_{\R_c}^\infty \rmd\R \frac{1}{\sqrt{2\pi} \sigma_\R} e^{-\frac{\R^2}{2\sigma_\R^2}}
    \approx \frac{\sqrt{2}\sigma_\R}{\sqrt{\pi}\R_c}e^{-\frac{\R_c^2}{2\sigma_\R^2}} \ .
\end{equation}
We build our inflationary potentials so that $\Omega_\text{PBH}$ calculated from the power spectrum in the Gaussian approximation matches the observed dark matter abundance today. We will see that the Gaussian approximation strongly underestimates the real abundance. 

\section{Results} \label{sec:res}

\subsection{Analytical toy potentials}

The details of how we discretise the continuum equations and solve them numerically are given in appendix \ref{sec:num}. We first test our procedure on two toy model potentials studied in \cite{Ezquiaga:2019ftu}, where $P(N)$ was calculated analytically. We consider the family of cubic potentials
\bea
  V(\phi) &=& V_0 [ 1 + \a (\phi-\phi_0) + \b (\phi-\phi_0)^3 ] \ ,
\eea
where $V_0>0$, $\a\geq0$, $\b\geq0$, and $\phi_0$ are constants. The potential is called flat or tilted if $\a = 0$ or $\a \neq 0$, respectively.

We use the simplified treatment of noise, where the momentum kicks vanish and the field kick amplitude is given by \re{noise_correlator_simplified}, for two reasons. First, we are interested in checking whether we can replicate the results of~\cite{Ezquiaga:2019ftu}, which used the simplified treatment. Second, our full treatment cannot start directly in the flat section of the potential, as modes need to be first evolved before any kicks can be given. The full treatment also requires the potentials to be classically traversable, which is not the case for some of the parameter values considered here.

\begin{figure}[t!]
\centering
\includegraphics{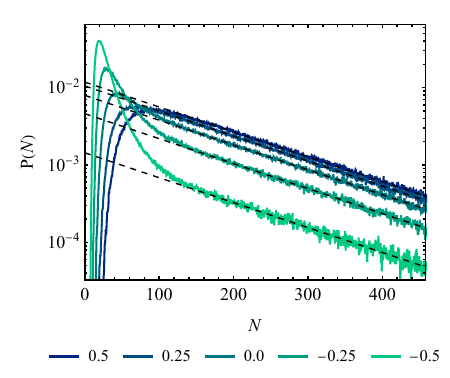}
\includegraphics{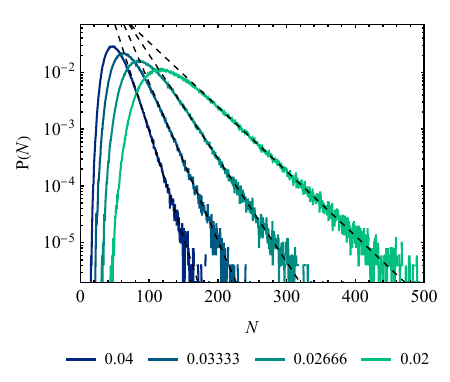}
\caption{Our results for the probability distribution $P(N)$ for the potentials used in \cite{Ezquiaga:2019ftu}. For the flat cubic potential (left), the different colours correspond to different values of $(\phi - \phi_0)/\Delta \phi_{\text{well}}$. For the tilted cubic potential (right), the different colours correspond to different values of the parameter $\alpha$. The black dashed lines are exponential fits to the numerical results.}
\label{fig:toy}
\end{figure}

In the flat cubic potential case we set $V_0 = \frac{6 \pi^2}{25}$, $\alpha = 0$, and $\beta = 0.01$, and vary the starting point of the field so that $(\phi - \phi_0)/\Delta \phi_{\text{well}}$ takes the values $-0.5$, $-0.25$, $0$, $0.25$, $0.5$, where $\phi_0 = \Delta \phi_{\text{well}}/2 + 0.3/\sqrt{\beta}$ and $\Delta \phi_{\text{well}} = 2 \left(150 \beta \right)^{-1/3}$. The results are shown in figure~\ref{fig:toy} (left). In all cases, the exponential fit $P(N)=Ae^{-BN}$ to the tail of the distribution gives robustly the slope $B = 7.4 \times 10^{-3}$. Our results are to be compared to those shown in figure 9 of \cite{Ezquiaga:2019ftu}. The analytical leading term flat approximation in \cite{Ezquiaga:2019ftu} gives the slope $B = 8 \times 10^{-3}$, while including higher order terms gives $B = 7.4 \times 10^{-3}$, in perfect agreement with our result\footnote{The analytical result for the slopes is only given to leading order in \cite{Ezquiaga:2019ftu}, but the numbers including higher order corrections have been privately communicated to us by the authors.\label{fn:privcomm}}.

In the tilted case we set $V_0 = \frac{3 \pi^2}{25}$, $\phi_0 = 0.1/\alpha$, $\beta = 10^{-3}$, and $\alpha$ takes the values $0.02$, $0.02666$, $0.03333$ and $0.04$. The results are shown in figure \ref{fig:toy}~(right). We again fit an exponential and find that the slope $B$ takes the values $0.026$, $0.042$, $0.064$ and $0.087$, respectively. These results are to be compared to those in figure 11 of \cite{Ezquiaga:2019ftu}. The leading term flat approximation in \cite{Ezquiaga:2019ftu} gives the corresponding slopes $0.022$, $0.037$, $0.057$ and $0.081$, while including higher order terms\footref{fn:privcomm} gives values $0.025$, $0.042$, $0.062$ and $0.086$, respectively, close to our results, especially given the errors shown in \fig{fig:toy}.

These results show that our algorithm works as intended, and we confirm the existence of the exponential tails. We now move to our realistic potentials.

\subsection{Realistic potentials} \label{sec:real}

\subsubsection{Potentials for the four cases}

The cubic models considered above describe potentials only around the region where the PBH seeds are generated. In a realistic model, we have to fit not only the PBH adundance, but also CMB observables, which correspond to larger inflaton field values. When building the models, the location of the USR region is determined by the relation \re{eq:M_PBH} between the PBH mass and wavenumber. For longer wavelengths, the Hubble volume at collapse is bigger, so the PBH mass is larger. Hence, the larger the PBH mass, the closer the USR region is to the CMB scales, and correspondingly the more difficult it is to have a potential that leads to USR and produces enough PBHs without spoiling the CMB predictions. Although the difference in the wavenumber of modes that collapse to form PBHs and modes in the observed CMB region can be exponentially large, the difference in field values is moderate. For the potentials we consider, the field value at the CMB pivot scale is between 3 to 6 times its value at the beginning of USR, which we define as the period when the second SR parameter $\eta=-\ddot{\bar\phi}/(H \dot{\bar\phi})$ is larger than 1.

\begin{figure}[t!]
\centering
\includegraphics{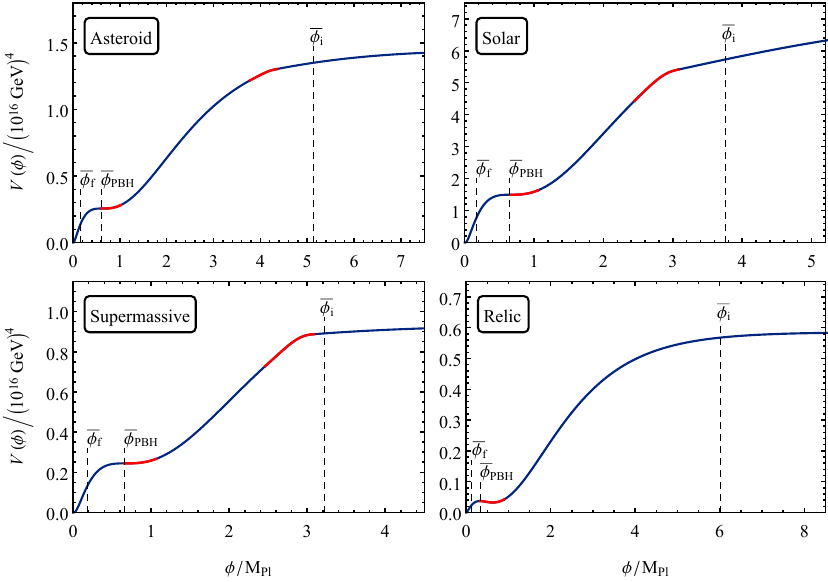}
\caption{The potentials for the asteroid, solar, supermassive, and relic case. The dashed lines mark the initial field value $\phil_i$ at the CMB scale, the final field value $\phil_f$ near the end of inflation, and the PBH scale $\phil_\text{PBH}$ corresponding to the Hubble exit of the last mode that gives a kick. The red sections indicate the USR region (small $\phi$) and the kink that interpolates between the low and high sections of the potential (large $\phi$).}
\label{fig:pot}
\end{figure}

We look at four cases: PBHs with asteroid mass, solar mass, and Planck mass, which can be dark matter, as well as PBHs with mass $1800M_\odot$, which can act as seeds for supermassive black holes~\cite{Carr:2020xqk, Green:2020jor, Carr:2020gox}. Our starting point for the inflaton potentials is non-minimally coupled Standard Model Higgs inflation~\cite{Bezrukov:2007ep, Rubio:2018ogq}, and we use loop corrections in the chiral Standard Model to create a shallow minimum that leads to USR \cite{Rasanen:2018fom} (see also \cite{Ezquiaga:2017fvi,Ballesteros:2017fsr,Kannike:2017bxn, Ballesteros:2017fsr, Bezrukov:2017dyv, Hertzberg:2017dkh}). This model can fit the PBH abundance and the CMB observables simultaneously only if the PBHs are Planck mass relics. In the other cases, we adjust the potentials by hand on the CMB scales, like we did in~\paperI. In practice, this means changing numerical factors in the potential above a threshold that lies between the PBH and CMB scales, as described in appendix \ref{sec:pot}. The resulting potentials for the four cases are shown in \fig{fig:pot}, and the masses, CMB observables, and PBH parameters are listed in table~\ref{tab:numbers}.

We use the metric formulation of general relativity. In \cite{Rasanen:2018fom}, PBHs from USR Higgs inflation were considered also in the Palatini formulation. This makes the potential much flatter, and as a consequence the PBH abundance is much more sensitive to the input parameters, and it is difficult to find the parameters for which the Gaussian approximation gives even roughly the observed PBH abundance.

\begin{table}
\begin{center}
\begin{tabular}{lcccc}
\toprule
& Asteroid & Solar & Supermassive & Relic \\
\midrule
$M$ & $7.2\times10^{-15} M_\odot$ & 4.7 $M_\odot$ & $1.8\times10^3 M_\odot$ & $1.4\times10^3$ kg \\
$n_s$ & 0.966 & 0.965 & 0.962 & 0.958 \\
$r$ & 0.01 & 0.05 & $8\times10^{-3}$ & $5\times10^{-3}$ \\
$\a_s$ & $-3\times10^{-4}$ & $7\times10^{-4}$ & $-7\times10^{-4}$ & $-9\times10^{-4}$ \\
\midrule
\textbf{Gaussian} &&&& \\
$\sigma_\R^\text{G}$ & 0.122 & 0.1739 & 0.118 & 0.149 \\
$\Omega_\text{PBH}^\text{G}$ & 0.13 & 0.17 & $1.4\times10^{-5}$ & 0.11 \\
\midrule
\textbf{Full} &&&& \\
$\sigma_\R$ & 0.123 & 0.1744 & 0.119 & 0.184 \\
$A$ & $1699\pm61$ & $1403\pm26$ & $2136\pm27$ & $225\pm5$ \\
$B$ & $32.7\pm1.2$ & $26.7\pm0.5$ & $41.3\pm0.5$ & $4.38\pm0.09$ \\
$\Omega_\text{PBH}^\text{fit}$ & $(1.6\pm0.4)\times10^4$ & $1.58\pm0.06$ & $0.049\pm0.01$ & $(2.38\pm0.06)\times10^7$ \\
$\Omega_\text{PBH}^\text{num}$ & $-$ & 1.55 & $-$ & $2.39 \times 10^7$ \\
$\Omega_\text{PBH}^\text{fit}/\Omega_\text{PBH}^\text{G}$ & $10^5$ & $10$ & $10^3$ & $10^8$ \\
\midrule
\textbf{Simplified} &&&&\\
$A$ & $2080\pm63$ & $998\pm4$ & $1833\pm58$ & $3995\pm57$ \\
$B$ & $40.0\pm1.2$ & $19.0\pm0.1$ & $35.4\pm1.1$ & $77.8\pm1.1$ \\
$\Omega_\text{PBH}$ & $26\pm12$ & $(1.25\pm0.01)\times10^2$ & $17\pm5$ & $(5.5\pm4.9)\times10^{-24}$ \\
$\Omega_\text{PBH}/\Omega_\text{PBH}^\text{G}$ & $10^2$ & $10^3$ & $10^6$ & $10^{-23}$ \\
\bottomrule
\end{tabular}
\end{center}
\caption{Masses, CMB observables and PBH parameters for the four potentials. Here $M$ is the initial PBH mass, $n_s$ is the CMB scalar spectral index, $r$ is the tensor-to-scalar ratio, and $\a_s$ is the running of the scalar spectral index. In the rows labeled ``Gaussian", $\sigma_\R^\text{G}$ is the standard deviation of the perturbations calculated from the power spectrum as explained above \eqref{eq:gaussian_beta_0}, and $\Omega_\text{PBH}^\text{G}$ is the corresponding PBH abundance today. In the rows labeled ``Full", $\sigma_\R$ is the standard deviation of the Gaussian fit to the full $P(N)$, $A$ and $B$ characterise the exponential fit $P(N)=e^{A-BN}$ to the tail, $\Omega_\text{PBH}^\text{fit}$ is the PBH density fraction today calculated from the exponential fit, and $\Omega_\text{PBH}^\text{num}$ is the density fraction calculated directly from $P(N)$ by numerical integration. The rows labeled ``Simplified" give the corresponding numbers for the simplified treatment of kicks \eqref{noise_correlator_simplified}. For relics, the loss of mass through evaporation has been taken into account in $\Omega_\text{PBH}$. The errors are based on a jackknife analysis with 20 subsamples.}
\label{tab:numbers}
\end{table}

\subsubsection{Asteroid mass dark matter}

In the mass window $\sim 10^{-15}$--$10^{-11} M_\odot$ PBHs can constitute all of the dark matter \cite{Carr:2020xqk, Green:2020jor, Carr:2020gox}. We consider a value close to the lower end of this range, $7.2\times10^{-15} M_\odot$, around the mass of the asteroid Eros. As discussed above, we adjust the renormalisation group improved Higgs potential by hand in the CMB region to make the spectral index less red, to fit observations. The tensor-to-scalar ratio is compatible with observations even without the adjustment.

\begin{figure}[t!]
\centering
\includegraphics{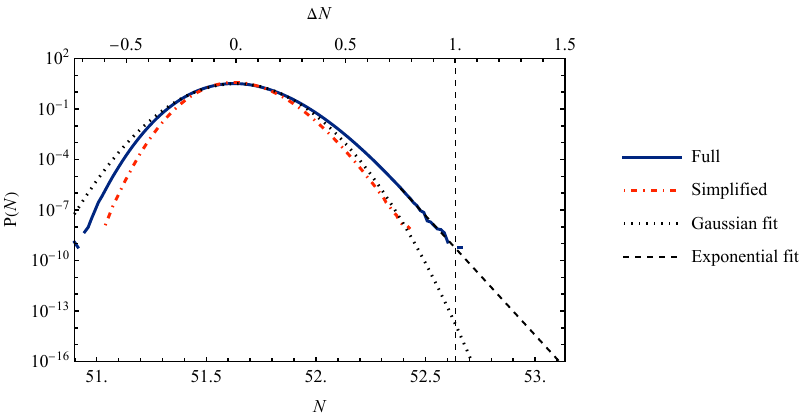}
\caption{The probability distribution $P(N)$ for the asteroid case. The bottom label gives the number of e-folds at the end of the simulation, the top label gives the deviation from the mean. The blue solid line is the full numerical result. The black dotted line is a Gaussian fit, and the black dashed line is the exponential fit to the tail. The red dash-dotted line is the simplified treatment. The vertical line marks the collapse threshold $\Delta N = 1$.}
\label{fig:asteroid}
\end{figure}

We have run $1024 \times 10^{8}$ simulations of the inflationary dynamics, each representing a realisation of the solution to (the discretised version of) the set of equations (\ref{Langevin}), \re{Langevin2}, (\ref{eq:background_H_eom}), and (\ref{eq:field_mode_eom_consistent}). The main results for this potential have already been reported in~\paperI. The probability distribution $P(N)$ is shown in \fig{fig:asteroid}, and some relevant numbers are given in \tab{tab:numbers}. For $|\Delta N|\gtrsim0.5$ the shape deviates significantly from a Gaussian, with an exponential tail. The distribution is clearly skewed towards large $\Delta N$. This can be easily understood. Each kick moves the field forward or back with equal probability, but in USR the field velocity decays rapidly.  Hence, the longer the field spends in the USR region, the slower it moves, so being kicked back increases $\Delta N$ more than being kicked forward decreases it. The effect is sizeable only in the tail of the distribution; the mean $\bar N$ remains essentially unchanged. We have data points up to $\Delta N=1.09$, but the region $\Delta N\gtrsim0.95$ is poorly resolved. The computational cost of resolving further into the tail would be prohibitive given our computational resources. We fit an exponential $e^{A-BN}$ to the tail in the range $\Delta N=0.75\ldots0.95$, where the statistics are robust. We then calculate the abundance from \re{beta} using an extrapolation of the fit. As shown in \tab{tab:numbers}, the PBH abundance is enhanced by a factor of $\sim 10^5$ compared to the Gaussian case.

To assess the importance of mode evolution in a background affected by previous kicks, we have done $256\times10^{8}$ runs with the simplified treatment where the momentum kicks vanish and the amplitude of the field kicks is given by \re{noise_correlator_simplified}, as discussed in \sec{sec:remarks}. This leads to a much steeper slope, as shown in figure \ref{fig:asteroid}. In this case, we only have points up to $\Delta N=0.79$. Fitting an exponential to the well-resolved range $\Delta N=0.6\ldots0.7$ and extrapolating to $\Delta N \geq 1$ gives $\Omega_\text{PBH}=26$. This is $\sim 10^2$ times higher than in the Gaussian case, but a factor $\sim 10^3$ smaller than in the full treatment. The result is subject to extrapolation errors, but it is clear that the simplified treatment significantly underestimates the stochastic enhancement of the PBH abundance.

We also considered a recursive treatment where we first evolve the mode functions on the FLRW background without noise, and then do a run where these modes are used to calculate the noise amplitude \re{noise_correlator}. The results are (within the statistical error) indistinguishable from those of the full treatment, unless $\sigma$ is so small that the kicks are pushed close to the end of inflation. While the simplified treatment above shows that it is crucial to take into account mode evolution when calculating the noise, the recursive treatment establishes that the change of the coarse-grained background due to noise has little impact on the mode evolution (except for small $\sigma$). This is perhaps not surprising, given that the Hubble rate, which characterises the background evolution, is almost constant (except during the transition from SR to USR), so the kicks don't actually perturb the background much. Thus, while the full treatment is not Markovian, and hence does not reduce to a Fokker--Planck equation in principle, the non-Markovian nature of the dynamics turns out to be irrelevant. Furthermore, we have checked that the difference in $\Omega_\text{PBH}$ is only $21\%$ (\ie smaller than our 25\% jackknife error) if we fix the amplitude of the kicks to the square root of the variance (which evolves in time), and randomise only their sign, \ie whether the field gets kicked forward or backward. This approximation may simplify analytical calculations of the stochastic effects.

\begin{figure}[t]
\centering
\includegraphics{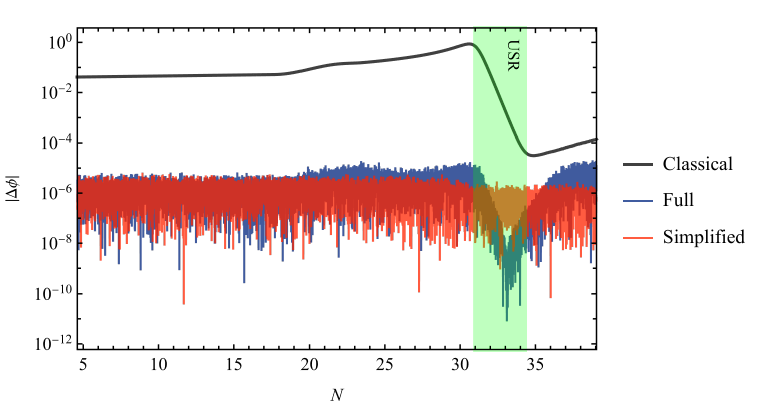}
\caption{Stochastic kicks as a function of e-folds for the asteroid case. The black solid line is the classical evolution of the field, the kicks in the full treatment are in blue, and the kicks in the simplified treatment are in red. The classical evolution has been scaled by $\dd{N}^{-1}$ and the kicks by $\dd{N}^{-1/2}$ to make the overall amplitudes in the plot independent of the time step $\dd{N}$. (See appendix \ref{sec:num} for details.) The green area denotes the USR region. The plot shows all the kicks in a typical run.}
\label{fig:asteroidkicks}
\end{figure}

\begin{figure}
\centering
\includegraphics{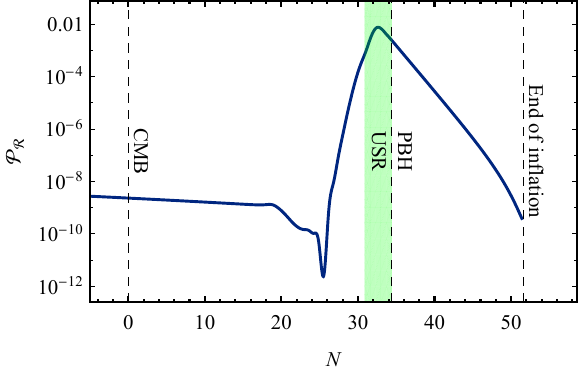}
\caption{The power spectrum $\PR(k)$ for the asteroid case over different scales, evaluated at the end of inflation when all the super-Hubble scales have frozen. The e-fold number $N$ corresponds to the time the mode exits the Hubble radius. The dashed lines mark the CMB scale and the end of inflation, and the green area indicates USR. The last scale that contributes to the PBHs, corresponding to the end of USR, is also marked.}
\label{fig:asteroidPR}
\end{figure}

In figure \ref{fig:asteroidkicks}, we show the kicks in the full and in the simplified treatment for a typical run. In the simplified treatment, the typical kick amplitude is almost constant. In contrast, in the full treatment the kick amplitude varies by orders of magnitude, with a distinct dip during USR. This structure traces back to the curvature power spectrum, shown in figure~\ref{fig:asteroidPR}, which drops for modes that exit the Hubble radius just before the start of USR due to a cancellation between terms in the transition region \cite{Byrnes:2018txb, Passaglia:2018ixg, Carrilho:2019oqg, Ozsoy:2019lyy, Tasinato:2020vdk}, and peaks for modes that exit the Hubble radius during USR. Note the offset between the two figures. The strongest perturbations originate from USR, but give their kicks later, when they exit the coarse-graining scale, by which time USR is already over. After the end of USR, the kick amplitude continues to rise, and the variance is a maximum of 80 times larger in the full treatment than in the simplified treatment. This should be compared to the classical drift of the field, which dips during USR when the field slows down, and grows again afterwards. In practice, the kicks have the strongest effect somewhat after the end of USR, when the kicks have already grown to large values but the classical drift is still small. Individual kicks are much smaller than the classical contribution, which may in part explain why fixing the kick amplitude to the square root of the variance gives results indistinguishable from the full treatment: many kicks are required to have an impact.

Decreasing the coarse-graining scale parameter $\sigma$ from $10^{-2}$ to $10^{-4}$ changes $\Omega_\text{PBH}$ by only 4\%, much less than our statistical error of 25\%. For $\sigma = 10^{-5}$ the abundance grows by a factor of 3, and for $10^{-6}$ it grows by a factor of 5. Making $\sigma$ smaller increases the offset between when the mode exits the Hubble radius and when it delivers its kick. As long as the kick amplitude remains similar, the ensuing difference in the results is small, but it grows as the kicks are moved closer to the end of inflation. The agreement between the full and recursive treatment also degrades in this case. The difference between the full and recursive treatment grows to an order of magnitude for $\sigma = 10^{-7}$. For $\sigma=10^{-8}$, the kicks are pushed to the very end of inflation, and their amplitude becomes large. The $P(N)$ distribution in the full treatment changes drastically, becoming similar to the relic case discussed below in \sec{sec:relic}. We then also encounter the issue that some of the simulations never finish (as also happens in the relic case). A lower limit for $\sigma$ has been given from various requirements, including that the noise is given by the simplified form \re{noise_correlator_simplified} \cite{Starobinsky:1994, Levasseur:2013tja}, a derivation of the stochastic formalism from quantum field theory \cite{Tokuda:2017fdh, Tokuda:2018eqs}, SR remaining an attractor \cite{Grain:2017dqa}, and the validity of neglecting heavier fields in the multifield case \cite{Pinol:2020cdp}. The requirement in \cite{Grain:2017dqa} that the slope of the potential should not change significantly between the time a mode exits the Hubble radius and gives its kick agrees with the condition we find for the results to be insensitive to the value of $\sigma$.

\subsubsection{Solar mass dark matter}

Observations seem to imply that PBHs around the solar mass cannot form all of the dark matter, although some of the constraints may be ameliorated by nonstandard clustering \cite{Carr:2020xqk, Green:2020jor, Carr:2020gox}. In any case, such PBHs could contribute a subdominant component of dark matter. We consider PBHs with mass $4.7M_\odot$. As in the asteroid case, the USR region that seeds PBHs is produced by renormalisation group running, and we adjust the CMB part of the potential by hand to make the spectral index less red to fit observations. The tensor-to-scalar ratio fits the CMB data available up to Sept.~2021 \cite{Akrami:2018odb} with or without the adjustment.\footnote{As this paper was being finalised, a new combined analysis of new BICEP/Keck, Planck and WMAP CMB data appeared \cite{BICEPKeck:2021gln}. The updated constraint $r<0.036$ excludes our solar mass potential, which has $r=0.05$. This could be fixed by adjusting the height of the potential in the same way as for the supermassive case discussed in \sec{sec:super}. However, this would require re-running all simulations for the solar mass case, which would take a lot of computational resources without changing the qualitative conclusions.}

\begin{figure}[t]
\centering
\includegraphics{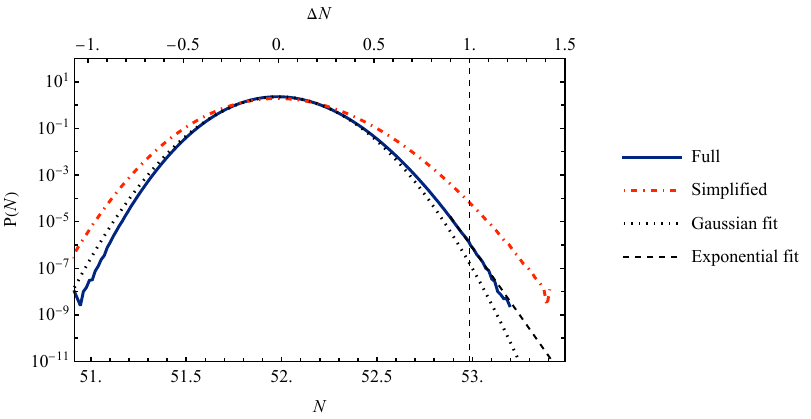}
\caption{The probability distribution $P(N)$ for the solar case, similarly to \fig{fig:asteroid}.}
\label{fig:solar}
\end{figure}

\begin{figure}[t!]
\centering
\includegraphics{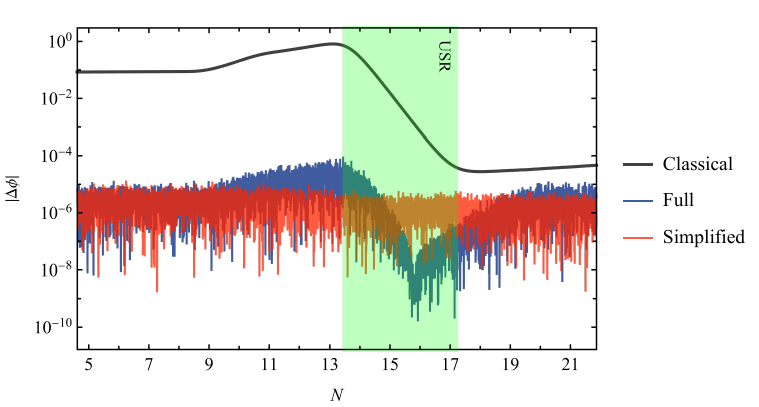}
\caption{Stochastic kicks for solar case, similarly to \fig{fig:asteroidkicks}.}
\label{fig:solarkicks}
\end{figure}

\begin{figure}
\centering
\includegraphics[scale=1]{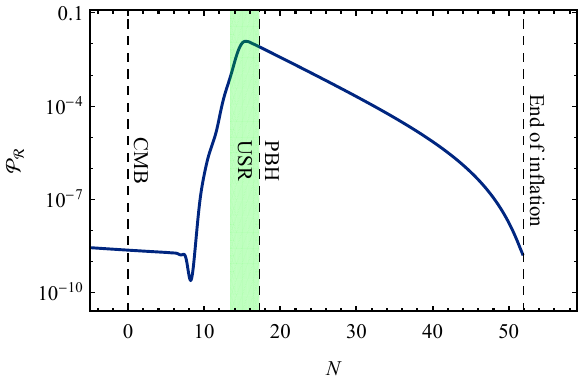}
\caption{The power spectrum $\PR(k)$ for the solar case, similarly to \fig{fig:asteroidPR}.}
\label{fig:solarPR}
\end{figure}

As the PBH mass is a factor of $\sim 10^{15}$ higher than in the asteroid case, fewer PBHs are needed today to match the observed dark matter abundance. However, as heavier PBHs collapse later, they have less time to increase their fraction of the energy density. As a result, the initial PBH abundance has to be larger than in the asteroid case, as quantified in \re{eq:omega_PBH}. Therefore, the variance from the power spectrum is larger, as shown in \tab{tab:numbers}. The slope of the full numerical $P(N)$ distribution shown in \fig{fig:solar} also turns out to be less steep than in the asteroid case. With $256\times10^{8}$ simulations, the distribution reaches up to $\Delta N=1.20$. The mean $\bar N$ is again unchanged from the Gaussian result. We can now numerically integrate the PBH abundance, with no need to extrapolate to reach the collapse threshold. The abundance is enhanced by one order of magnitude compared to the Gaussian case. The tail is well fit by an exponential, which gives the same result for the abundance as numerical integration of the distribution.

In the simplified treatment, with $256\times10^{8}$ simulations we reach up to $\Delta N=1.45$. In contrast to the asteroid case, now the simplified treatment gives a larger enhancement than the full treatment, by two orders of magnitude. The enhancement over the Gaussian case is three orders of magnitude. In \fig{fig:solarkicks}, we show the kicks for a typical run in the full and simplified treatments. The maximum amplitude of the kicks after USR in the full treatment is a factor of 5 larger than in the simplified treatment; the difference is much smaller than in the asteroid case. This enhancement after USR in the full treatment is not sufficient to overcome the dip during USR, compared to the steady kicks in the simplified case, which explains why the simplified case gives overall a larger enhancement. As in the asteroid case, the recursive treatment gives results indistinguishable from the full treatment after one iteration.

The calculation is less sensitive to the choice of the coarse-graining scale than in the asteroid case. Decreasing $\sigma$ from $10^{-2}$ to $10^{-8}$ changes $\Omega_\text{PBH}$ by less than 30\%. The effect is large only when the kicks are pushed to the very end of inflation, which corresponds to the value $\sigma=10^{-16}$. 

As can be seen from \fig{fig:solarPR}, the USR region is now closer to the CMB region than in the asteroid case: because the PBHs are more massive, their seeds have longer wavelengths and exit the Hubble radius earlier. In the asteroid case, USR begins 31 e-folds after the CMB pivot scale exits the Hubble radius, but in the solar case this interval is 14 e-folds. In terms of field values, the difference is less drastic: the value of the field changes by $4\Mpl$ between the CMB pivot scale exit and the start of USR in the asteroid case and by $3\Mpl$ in the solar case. The larger the mass, the smaller the separation, and the more difficult it is to keep the CMB perturbations unaffected by the USR region. This is even clearer in the case of supermassive PBHs.

\subsubsection{Supermassive black hole seeds} \label{sec:super}

Apart from dark matter, PBHs have been suggested as seeds for supermassive black holes observed at high redshifts. It is not clear how these objects can form sufficiently rapidly in the early universe \cite{Carr:2020xqk, Carr:2020gox}. There may be an astrophysical explanation, but initial PBHs that jumpstart the accretion process are also a possible solution. We consider initial seeds with mass $1.8\times10^3 M_\odot$, which then grow by accretion to $10^9 M_\odot$, with abundance today $\Omega_\text{PBH}=1.4\times10^{-5}$ in the Gaussian approximation. USR begins 10 e-folds after the exit of the CMB pivot scale, which is only a difference of $2\Mpl$ in the field value. In addition to adjusting the potential in the CMB region by hand, as in the previous cases, we now have to also adjust the overall height of the whole potential. Otherwise, the tensor-to-scalar ratio $r$ grows with the PBH mass as the USR and CMB regions come closer, and exceeds the observational CMB upper limit $0.036$ \cite{BICEPKeck:2021gln}, or even the previous one $r<0.061$ \cite{Akrami:2018odb}.\footnote{In the Palatini formulation of general relativity mentioned above, $r$ is suppressed \cite{Bauer:2008, Rasanen:2018fom}, so there would be no such problem. However, the spectral index would still need to be adjusted by hand.}

\begin{figure}[t!]
\centering
\includegraphics{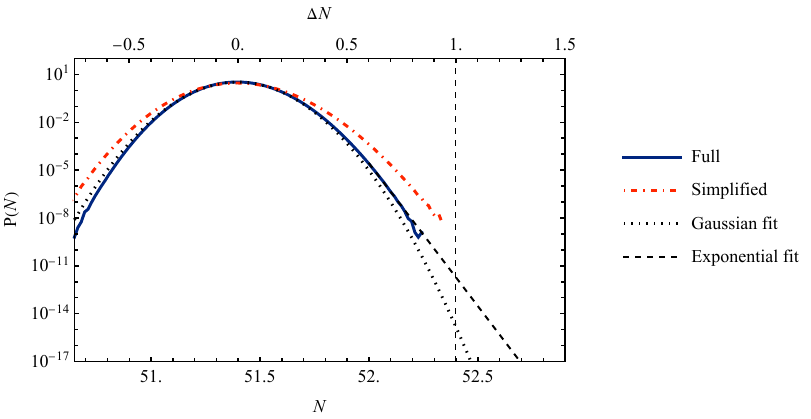}
\caption{The probability distribution $P(N)$ for the supermassive case, similarly to \fig{fig:asteroid}.}
\label{fig:supermassive}
\end{figure}

\begin{figure}[t!]
\centering
\includegraphics{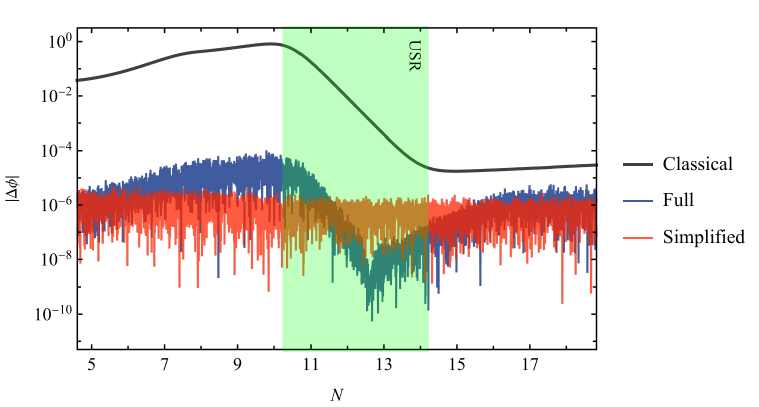}
\caption{Stochastic kicks as a function of e-folds for the supermassive case, similarly to \fig{fig:asteroidkicks}.}
\label{fig:supermassivekicks}
\end{figure}

\begin{figure}
\centering
\includegraphics[scale=1]{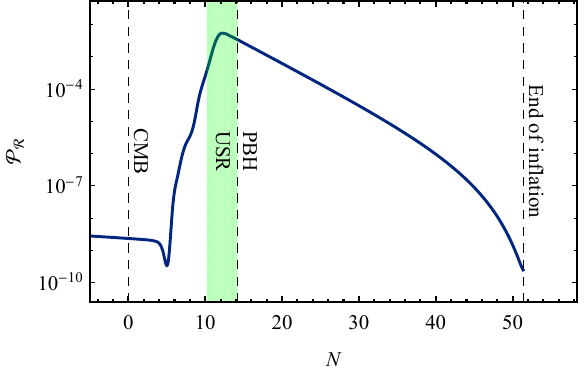}
\caption{The power spectrum $\PR(k)$ for the supermassive case, similarly to \fig{fig:asteroidPR}.}
\label{fig:seedPR}
\end{figure}

With $1024\times10^{8}$ simulations, the $P(N)$ distribution for this case reaches up to $\Delta N=0.83$, and is shown in \fig{fig:supermassive}. The mean $\bar N$ is again unchanged from the Gaussian result. Fitting an exponential to the well-resolved range $0.6\ldots0.75$ and extrapolating beyond $\Delta N=1$ gives an enhancement of $10^3$ over the Gaussian case. In the simplified case, $256\times10^{8}$ simulations reach up to $\Delta N=0.93$, leading to an extrapolated enhancement of $\sim 10^6$ over the Gaussian case. As in the solar case, the simplified treatment overestimates the enhancement of the PBH abundance. We show typical kicks in \fig{fig:supermassivekicks}. As in the solar case, the maximum amplitude of the kicks after USR is only a factor $5$ larger in the full treatment compared to the simplified case. Again, the recursive treatment gives results identical to the full treatment with one iteration.

As in the previous cases, the results depend very weakly on the value of $\sigma$: decreasing it from $10^{-2}$ to $10^{-8}$ changes $\Omega_\text{PBH}$ only by 70\%. Significant change is seen only at $\sigma = 10^{-17}$, when the kicks are delayed until the end of inflation.

In this model, the peak of the power spectrum (shown in \fig{fig:seedPR}), $\PR = 0.005$, is located at $k=5400$ Mpc$^{-1}$, close enough to the CMB scale that it can be probed by future spectral distortion observations \cite{Chluba:2019kpb, Chluba:2019nxa}. The model may already be in tension with the FIRAS measurements \cite{Mather:1993ij, Fixsen:1996nj, Byrnes:2018txb, Green:2020jor}, although a full spectral distortion analysis would be needed to draw definite conclusions. This is an important consideration for all USR models that generate PBHs in this mass range.

\subsubsection{Planck mass relic dark matter} \label{sec:relic}

There are essentially no observational constraints (apart from the dark matter relic abundance) on the scenario where dark matter consists of uncharged Planck mass objects (see \cite{Carney:2019pza} for a proposal of how to observe them). We consider PBHs that form with mass $1.4\times10^3$ kg and evaporate down to the Planck mass $\Mpl=4.3\mu$g. The usual semiclassical computation of Hawking evaporation breaks down close to the Planck mass, and it is possible that evaporation stops and leaves a stable relic \cite{MacGibbon:1987my, Barrow:1992hq, Carr:1994ar, Green:1997sz, Alexeyev:2002tg, Chen:2002tu, Barrau:2003xp, Chen:2004ft, Nozari:2005ah}. In the relic case, the power spectrum peaks on scales that leave the Hubble radius a little after the end of USR. For this reason, we have moved the PBH scale forward from the end of USR to capture the modes with the largest amplitude, as shown in \fig{fig:relicPR}. Unlike in the previous cases, there is no need to adjust the potential by hand. The non-minimally coupled Standard Model Higgs potential with renormalisation group corrections can produce both PBHs with the observed dark matter abundance (in the Gaussian approximation) and the observed CMB perturbations~\cite{Rasanen:2018fom}. As in the other cases, the feature that leads to USR is extremely sensitive to the parameters entering the renormalisation group running.

In the potentials shown in \fig{fig:pot}, a kink in the non-relic cases marks where the adjusted inflationary plateau is matched to the part of the potential that has the USR feature, whereas in the relic case, the potential continues smoothly from the CMB scale down to the USR region. In all four mass cases the USR region (unlike the CMB region) is roughly at the same field value: $\phil$ at the start of USR varies from $1.06\Mpl$ in the supermassive case to $0.89\Mpl$ in the relic case, although the difference in e-folds (and consequently PBH mass) is large. In the relic case, USR ends only about 2 e-folds before the end of inflation. This is why no adjustment is needed: the quantum corrections modify the successful Higgs inflation scenario only at the very end of inflation. However, this leads to various problems.

In the relic case, most modes that exit during USR do not have enough time to reach the coarse-graining scale before the end of inflation, unless we increase the value of $\sigma$ considerably compared to the other cases we have explored. The lack of a long SR period after USR also means that the perturbations $\R$ and $\zeta$ do not necessarily have time to equalise, which should be taken into account when relating the inflationary curvature perturbation to the post-inflationary density contrast. Furthermore, as the modes exit close to the end of inflation, they quickly re-enter the Hubble radius. Therefore PBHs may collapse during preheating instead of the radiation-dominated era as we have assumed. This might affect the collapse process and the evolution of their density fraction after formation, depending on the equation of state. The duration of preheating is also crucial here, and depends on the detailed shape of the potential and the couplings of the fields (including between the different components of the Standard Model Higgs doublet). In tree-level Higgs inflation (in the metric formulation), it is not clear whether preheating takes few e-folds or is almost instant \cite{Bezrukov:2008ut, GarciaBellido:2008ab, Figueroa:2009, Figueroa:2015, Repond:2016, Ema:2016, DeCross:2016, Sfakianakis:2018lzf, Hamada:2020kuy}. In the Palatini formulation, the field re-enters the very flat inflationary plateau during preheating, leading to new periods of inflation, and preheating takes less than one e-fold \cite{Rubio:2019, Tomberg:2021bll}. In general, if the inflaton remains close to its potential minimum during preheating and the potential is monomial, $V(\phi) \propto |\phi|^p$ (with $p\geq2$), preheating is relatively rapid, taking one to a few e-folds \cite{Antusch:2020iyq}, depending on the couplings. It therefore depends on the details of preheating whether the universe has had time to transition to radiation domination before the PBHs form.

Another problem is that in the transition from SR to USR the field velocity grows so much that inflation ends for 0.3 e-folds. (This problem seems to be specific to the potential; for example, with the Palatini formulation discussed above, similar relic potentials do not have this problem.) Then some modes that have become part of the background start crossing back inside the coarse-graining radius, and when inflation resumes, they exit for the second time. Our heuristic treatment of the stochastic formalism does not prescribe how these modes should be treated. Should they contribute stochastic kicks when they first exit the coarse-graining radius, and again at second exit? Such questions go beyond the simple system-environment split we use, and should be resolved by a proper derivation from quantum field theory. We opt to skip the second kicks, so we have a kick-free period until all modes that re-entered the coarse-graining radius have re-exited, after which the kicks resume. While this raises questions of principle, in practice these kicks are so subdominant that it makes little quantitative difference how we treat them. As shown in figure \ref{fig:relickicks}, the largest kicks during this period are roughly four orders of magnitude smaller than the classical evolution. Towards the end of inflation, the classical contribution increases as the background velocity $\pil$ starts to rise. The kicks also grow proportionally to $\pil$, so their impact relative to the classical contribution does not increase towards the end. Similar but less drastic growth of the background and kicks is also seen in the other mass cases.

In order for the modes relevant to Planck mass relics to have enough time to give their kicks before the end of inflation, we set $\sigma=0.35$, so that the delay between Hubble exit and giving kicks is short. The side effect of choosing such a large value of $\sigma$ is that while the modes around the USR region do classicalise in time according to our criterion, $\cosh(2r_k)>100$, the vast majority of modes before the USR region do not. Therefore, we miss out on about 42 e-folds worth of kicks starting from the beginning of the simulation until the first classicalised mode crosses the coarse-graining scale. Using such a large value of $\sigma$ can be naively expected to significantly overestimate the stochastic kicks, and not to give a sufficient decoupling between the system and the environment. In any case, as we have discussed, the treatment of stochastic effects in USR should be put on more firm footing before any reliable conclusions can be drawn.

\begin{figure}[t!]
\centering
\includegraphics{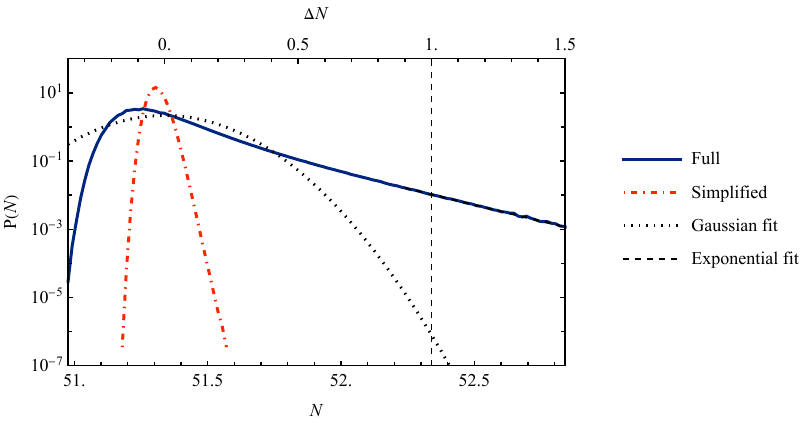}
\caption{The probability distribution $P(N)$ for the relic case, similarly to \fig{fig:asteroid}.}
\label{fig:relic}
\end{figure}

\begin{figure}[t!]
\centering
\includegraphics{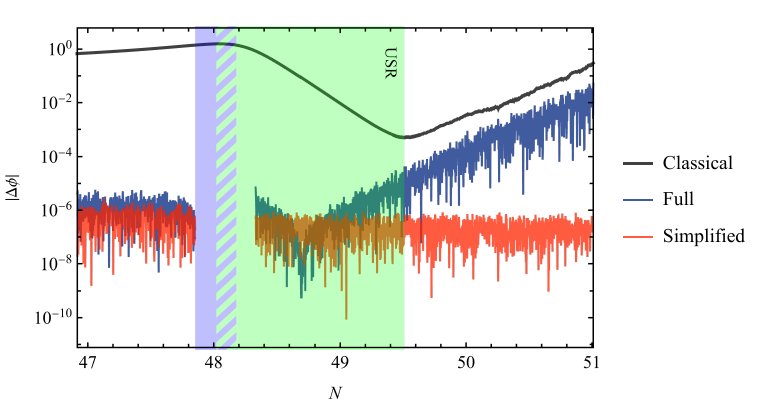}
\caption{Stochastic kicks as a function of e-folds for the relic case, similarly to \fig{fig:asteroidkicks}. The region where inflation temporarily ends is marked in violet, and its overlap with the USR region is indicated by the diagonal lines.}
\label{fig:relickicks}
\end{figure}

\begin{figure}[t!]
\centering
\includegraphics[scale=1]{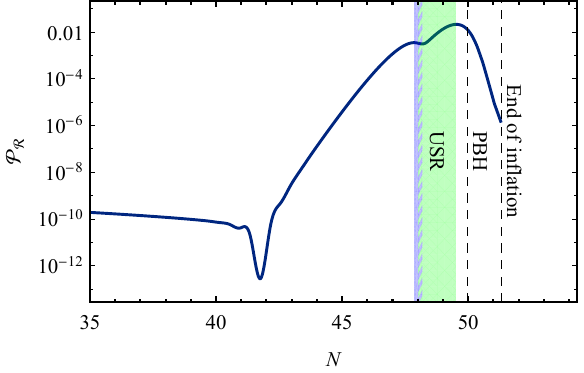}
\caption{The power spectrum $\PR(k)$ for the relic case, similarly to \fig{fig:asteroidPR}. The region where inflation temporarily ends is marked in violet, and its overlap with the USR region is indicated by the diagonal lines. Note that near this region, the same modes are depicted multiple times as they re-enter and again exit the Hubble scale. This leads to the dip in $\PR$ as a function of $N$ near $N=48$. The PBH scale is located somewhat after the end of USR to capture the peak in $\PR(k)$.}
\label{fig:relicPR}
\end{figure}

Running $256\times10^5$ simulations, the $P(N)$ distribution goes up to $\Delta N=3.77$. Figure \ref{fig:relic} shows that the changes due to the stochastic effects are qualitatively similar to the previous three cases, but more extreme. The kicks now distort the distribution everywhere, not only in the tail. The tail is still well described by an exponential. The mean value is barely affected: classical FLRW evolution has $\bar N=51.31$, and the full stochastic treatment gives $\bar N=51.34$. However, the distribution around the mean is no longer described by a Gaussian. As in the other cases, the first iteration of the recursive treatment gives results indistinguishable from the full calculation.

Numerical integration gives a PBH abundance that is enhanced by a factor of $10^8$ over the Gaussian calculation. The exponential fit gives results in agreement. However, the wide $P(N)$ distribution leads to serious interpretation problems. In approximately~6\% of the runs inflation does not end by $N=70$, when we stop the simulation, if the field has not reached the reached the end-of-simulation value $\bar{\phi}_{\text{f}}$. Instead, the field gets stuck in the local minimum. The corresponding points would be far to the right of the points plotted in \fig{fig:relic}. As we do not know how long the runs would continue, the shape of this far tail is unknown. The estimate of PBH abundance is thus not reliable, as the number of these unfinished runs is larger than the number of finished runs with $\Delta N\geq1$. We could also have the problem of eternal inflation, \ie diverging expectation value of the volume. Any exponential tail with an asymptotic slope $P(N)\propto e^{-BN}$ where $B\leq3$ leads to this problem. Although the value $B=4.38\pm0.09$ we find is safely above this limit, we cannot draw definite conclusions due to the runs that did not finish, which follow a different distribution. There is a related problem for any infinitely extended exponential tail. For example, for $3<B\leq6$ (including our relic case), the expectation value of the volume is finite, but the variance diverges, so the expectation value does not describe a typical member of the ensemble. Iterating this reasoning seems to lead to a problem of interpretation for all $P(N)$ with any exponential (or less steep) tail.

In the simplified treatment, the distribution is much sharper than the linear theory Gaussian. Running $256\times10^7$ simulations reaches only up to $\Delta N=0.26$. Such a drastic narrowing of the classical distribution by stochastic effects is not unknown. In \cite{Martin:2011}, where the same simplified treatment is used, stochastic effects significantly narrow $P(N)$, and reduce $\bar N$ by an order of magnitude. For the potential studied there, the field classically spends many e-folds in a flat region, while the noise kicks it out fast. In our case the number of e-folds in the USR region is classically small. The difference between the full and simplified treatment is due to the difference in the kick amplitudes shown in \fig{fig:relickicks}. (The interval without kicks corresponds to the temporary end of inflation.) The kicks in the full case are at their maximum four orders of magnitude larger than in the simplified case. Recall that we define USR as the time when the second SR parameter $\eta=-\ddot{\bar\phi}/(H \dot{\bar\phi})$ is larger than 1, so for some time after the transition from SR to USR, we have USR without inflation. In the simplified case all the runs finish.

Perhaps surprisingly, the results of the full treatment again depend only weakly on the value of $\sigma$: decreasing it to $10^{-8}$ changes $\Omega_\text{PBH}$ only by 70\%. Curiously, while we lose the peak modes by considering such a small $\sigma$, it also enables earlier modes (which get enhanced as we approach the end of inflation) to become classicalised and give their kicks close to the end of inflation instead, giving almost the same result. However, the fraction of simulations that finish is strongly affected. For instance, for $\sigma = 10^{-4}$, only 10\% of the simulations complete, in the rest the field gets stuck in the minimum of the potential. There does not appear to be a clear pattern to the relationship between this fraction and the value of $\sigma$.

In the relic case, our calculation also has a problem with gauge-dependence. In contrast to the other cases, the effect of the transformation from the spatially flat gauge to the uniform $N$ gauge is above 10\% even away from the narrow spike mentioned in \sec{sec:remarks}. It would be straightforward to take into account the gauge transformation, but given the other problems with the relic case, there is little point in correcting only this issue. Furthermore, as discussed above, a proper derivation of the Langevin equations from first principles, including a full accounting of the gauge issue, remains to be done. 

In summary, for the relic case we simply present the data obtained from our simulations, acknowledging that no robust conclusion about the dark matter relic abundance can be reached without a more detailed treatment.

\section{Conclusions} \label{sec:conc}

Using the $\Delta N$ formalism, we have calculated numerically the probability distribution $P(N)$ of e-folds of inflation in order to find the stochastic correction to the abundance of primordial black hole (PBH) seeds produced during ultra-slow-roll (USR). We have used the same set of stochastic equations as in \paperI~\cite{Figueroa:2020jkf}, which for the first time consistently took into account both the dependence of the noise that drives the coarse-grained background on the evolution of the perturbations, and the change in the evolution of the perturbations due to the change in the background. This leads to a non-Markovian process.

We test our algorithm by considering USR in some toy model potentials used in \cite{Ezquiaga:2019ftu}, finding results in close agreement. We then build realistic inflationary potentials using as a starting point Higgs inflation~\cite{Bezrukov:2007ep, Rubio:2018ogq}, where loop corrections in the chiral Standard Model provide a feature at the desired PBH scale \cite{Rasanen:2018fom}. We have adjusted these potentials by hand (except in the relic case) in the CMB region to give the right spectral index. In the supermassive case, we have also adjusted the overall height of the potential to give a sufficiently small tensor-to-scalar ratio. (This would also be needed in the solar case to fit the latest CMB data \cite{BICEPKeck:2021gln}.) We have considered four cases, tuning the potentials to have a shallow minimum that leads to USR. This leads to the generation of roughly the observed abundance of dark matter in the form of PBHs of asteroid, solar, or Planck mass, or alternatively the seeds of supermassive black holes with density parameter $10^{-5}$ today, in the Gaussian approximation. The stochastic effects generate an exponential tail for $P(N)$, which dominates over the linear theory Gaussian~\cite{Pattison:2017mbe, Ezquiaga:2019ftu, Figueroa:2020jkf}. The stochastic noise enhances the PBH abundance by roughly 5, 1, 8, and 3 orders of magnitude in the asteroid, solar, relic, and supermassive case, respectively.

We compare our results to the usual treatment of stochastic inflation, where the modes evolve on the FLRW background, and the noise amplitude is fixed to its slow-roll super-Hubble limit, where it is given by the FLRW Hubble rate. We also compare to a recursive treatment where the modes first evolve on the FLRW background, and are then used as the noise for the next iteration \cite{Levasseur:2013ffa, Levasseur:2013tja, Levasseur:2014ska}. The recursive treatment gives the same results as our full treatment with just one iteration, while the simplified treatment significantly over-predicts or under-predicts the stochastic effect. In the simplified treatment, the enhancement over the Gaussian case is 2, 3 and 6 orders of magnitude in the asteroid, solar and supermassive cases, respectively. In the relic case, the simplified treatment predicts a negligible PBH abundance. This shows that the evolution of the modes that produce the stochastic noise has to be taken into account, whereas the change in the coarse-grained background in the evolution of perturbations is not important, at least for the potentials we consider. In other words, the non-Markovian nature of the process is unimportant.

We set the coarse-graining scale by the requirement that the mode squeezing parameter $r_k$ satisfies $\cosh(2r_k)>100$ for all modes when they give their kicks, so that they are sufficiently classical. The coarse-graining scale gives a delay between the time when the modes exit the Hubble radius and the time they join the background and give a kick to the evolution. The results are rather insensitive to the choice of the coarse-graining scale, as long as the delay is not so long that the amplitude of the kicks has changed significantly between Hubble exit and coarse-graining.

We have also considered the gauge-dependence of our results. We have verified numerically that in the asteroid, solar and supermassive cases the correction from the gauge transformation between the spatially flat gauge to the uniform $N$ gauge, computed as the ratio of the gauge correction over the mode amplitude, is less than $10^{-5}$ for most modes. There is a narrow spike of modes for which the gauge correction rises to a maximum of $\sim 10^{-2}$. These correspond to modes with very small amplitudes at the coarse-graining scale, and their contribution to the overall dynamics is small. 

In the case of Planck mass relics the gauge transformations have a large effect, and this model also suffers from other problems. Hence, no robust conclusions may be drawn. In this case, USR occurs right before the end of inflation, which leads to several issues. If we wanted each mode in the simulation to classicalise before crossing the coarse-graining scale, the modes that contribute most to PBH formation would not have enough time to give their kicks. We therefore set a lower coarse-graining scale, meaning we miss most kicks from the beginning of the simulation in order to preserve the ones around the USR region. The effect of the kicks is large, as they grow towards the end, and in $\sim$ 6\% of the runs they slow down the field so much that it gets stuck in the shallow minimum in the USR region, and inflation does not end by the time we finish the simulation. This percentage is larger than the fraction of simulations where $\Delta N$ exceeds the PBH formation threshold, so our calculation of the abundance is not reliable. There is also the problem that inflation temporarily ends before and during USR, and modes start crossing back through the coarse-graining scale. It is not clear how to treat this in the stochastic formalism. Furthermore, as the period of slow-roll after USR is very short, the relation between the curvature perturbations and the density contrast may not correspond to the usual one. Also, the assumption that PBHs form during radiation domination may need to be corrected, as the modes re-enter the Hubble radius quickly after inflation, so the collapse may happen during preheating.

Our results show that in models with USR, it is crucial to check the impact of stochastic effects on the curvature perturbation when considering extreme perturbations such as PBH seeds. They also indicate that fixing the noise amplitude to its slow-roll super-Hubble limit can be very inaccurate. The quantitative impact depends on the details of the potential, and has to be checked on a case by case basis. On a more technical side, to obtain the full treatment results for our four potentials, we have run over 2 million CPU hours of simulations in total. The need for such a considerable amount of computational time may hinder doing more similar calculations. To reduce the time needed to obtain $P(N)$, it is more practical to use the recursive procedure, where the mode functions are only evolved once, and then used as input for the noise, without having to evolve the modes again. The results of this recursive treatment for all the four potentials we have considered is indistinguishable (within the statistical errors) from those of the full treatment, but the running time is between $1\%$ to $10\%$ of that in the full treatment, depending on the potential.

As a final note, we remark that the derivation of the stochastic equations we have used is heuristic, and would need to be worked out from first principles in quantum field theory. The existing  derivations that lead to explicit Langevin equations assume slow-roll, and do not necessarily apply in USR.

\acknowledgments

We thank Peter Johansson and Vincent Vennin, Jose Mar\'ia Ezquiaga and Juan Garc\'ia-Bellido for discussion and correspondence. DGF (ORCID 0000-0002-4005-8915) is supported by a Ram\'on y Cajal contract with Ref.~RYC-2017-23493. This work was supported by the Generalitat Valenciana grant PROMETEO/2021/083, the Ministerio de Ciencia e Innovaci\'on grant PID2020-113644GB-I00, the Estonian Research Council grants PRG803, PRG1055, and MOBTT5, and by the EU through the European Regional Development Fund CoE program TK133 ``The Dark Side of the Universe''. This work used computational resources provided by the Finnish Grid and Cloud Infrastructure (urn:nbn:fi:research-infras-2016072533). The authors wish to acknowledge CSC - IT Center for Science, Finland, for computational resources.

\appendix

\section{Inflaton potential} \label{sec:pot}

The inflaton potentials used in this work come from the quantum-corrected potential of Higgs inflation in the Einstein frame, as described in detail in \cite{Rasanen:2018fom}. In summary, the potential can be written as
\begin{equation}
    V[\phi(h)] = c_0 \frac{\lambda_\mathrm{eff}(h)}{4}F(h)^4 \ ,
\end{equation}
where $\phi$ is the Einstein frame canonical field. It is related to the original Jordan frame Higgs field $h$ through the field transformation
\begin{equation}
    \frac{\de \phi}{\de h} = \frac{\sqrt{1+\xi h^2 + 6\xi^2 h^2}}{1+\xi h^2} \ ,
\end{equation}
where $\xi$ is a non-minimal coupling to gravity. The constant $c_0$ is an extra scaling factor introduced by hand for the supermassive case to reduce the tensor-to-scalar ratio $r$. The effective coupling $\lambda_\mathrm{eff}(h)$ contains the one-loop Coleman--Weinberg corrections evaluated so that all couplings run according to the one-loop beta functions of the chiral Standard Model. The function $F(h)$ is
\begin{equation}
    F(h) \equiv
\begin{cases}
    \frac{h}{\sqrt{1+\xi h^2}} \, , & h\leq h_1 \\
    \text{interpolation} \, , & h_1\leq h\leq h_2 \\ 
    c_1 \frac{h+c_2h_0}{\sqrt{c_3+\xi (h+c_2h_0)^2}} \, , & h \geq h_2 \ ,
\end{cases}
\end{equation}
where the first expression is the same as in standard Higgs inflation and is used near the PBH scale where $h<h_1$, and the third expression is modified by the coefficients $c_1,c_2,c_3$, so that it produces perturbations consistent with the observations in the CMB region where $h>h_2$. The reference scale $h_0$ is the location of the feature in the potential. The interpolation is done with an order five polynomial, chosen so that $F$ and its first and second derivatives are continuous everywhere. The interpolation gives the kinks in the potentials shown in \fig{fig:pot}.

The couplings featured in the quantum corrections are tuned so that the first derivative of the potential vanishes at $h_0$ and the second derivative is fixed by the second SR parameter $\eta=\frac{1}{V(\phi)}\frac{\de^2 V(\phi)}{\de\phi^2}$. In table~\ref{tab:potential_parameters} we give $h_0$ in terms of $\delta_0\equiv\frac{1}{\xi h_0^2}$, the value of $\eta$ at $h=h_0$, and the parameters in $F(h)$ for each of our example potentials.

In our numerical calculations, we use the full renormalisation group improved expression for $\lambda_\mathrm{eff}(h)$. It can be roughly approximated in the different cases as\footnote{Previous versions of the paper had a typo with $F(h)/F(h_0)$ inside the logarithms instead of $\sqrt{\xi}F(h)$.}
\begin{equation}
\begin{aligned}
\lambda_\mathrm{eff}(h) &\approx 2.556\times10^{-6} + 1.029\times10^{-5} \ln [\sqrt{\xi}F(h)] + 2.326\times10^{-5} \ln^2 [\sqrt{\xi}F(h)] &\;& \text{(asteroid)} \\
\lambda_\mathrm{eff}(h) &\approx 2.457\times10^{-6} + 9.799\times10^{-6} \ln [\sqrt{\xi}F(h)] + 2.316\times10^{-5} \ln^2 [\sqrt{\xi}F(h)]&\;& \text{(solar)} \\
\lambda_\mathrm{eff}(h) &\approx 2.464\times10^{-6} + 9.829\times10^{-6} \ln [\sqrt{\xi}F(h)] + 2.312\times10^{-5} \ln^2 [\sqrt{\xi}F(h)] &\;& \text{(supermassive)} \\
\lambda_\mathrm{eff}(h) &\approx 3.795\times10^{-6} + 1.1617\times10^{-5} \ln [\sqrt{\xi}F(h)] + 2.329\times10^{-5} \ln^2 [\sqrt{\xi}F(h)] &\;& \text{(relic)} \ .
\end{aligned}
\end{equation}

\begin{table}
\begin{center}
\begin{tabular}{lcccc}
\toprule
& Asteroid & Solar & Supermassive & Relic \\
\midrule
$\delta_0$ & 1.48137 & 1.484794 & 1.4878 & 1.5167 \\
$\xi$ & 38.8 & 16.332(54668015058) & 12.75 & 75 \\
$c_0$ & 1 & 1 & 0.1 & 1 \\
$c_1$ & 0.998 & 0.985 & 0.958 & -- \\
$c_2$ & 4.7 & 6.2 & 1.469 & -- \\
$c_3$ & 2.5 & 5 & 0.28 & -- \\
$h_1/h_0$ & 5.5 & 3 & 3 & $\infty$ \\
$h_2/h_0$ & 7 & 4 & 4 & -- \\
$\eta$ & 0.4141(0445743324537) & 0.141 & 0.1466 & 7.28023 \\
\bottomrule
\end{tabular}
\end{center}
\caption{The parameter values used to build our example potentials. In the relic case, the expression for the potential at $h<h_1$ applies for all $h$, since there is no need to adjust the CMB scales. The overall scale factor $c_0$ is only used for the supermassive black hole seeds. For the parameters with many digits, the full accuracy is needed to produce our results reliably; ignoring the digits in the parentheses gives Gaussian PBH abundances to within $10\%$.}
\label{tab:potential_parameters}
\end{table}

\section{Numerical implementation} \label{sec:num}

In our discretisation procedure, a single time step consists of the classical evolution of all quantities and a stochastic kick, where the kick is computed using the newly evolved values. The continuum equations of motion for the background and the perturbations read
\begingroup
\allowdisplaybreaks
\begin{align}
\tilde{\phi}' &= \tilde{\pi} \\
\tilde{\pi}' &= \left( \frac{1}{2}\tilde{\pi}^2 - 3 \right) \left(\tilde{\pi} + \frac{\tilde{V}_{,\phi}}{\tilde{V}} \right) \equiv f(\tilde{\phi},\tilde{\pi}) \\
\delta{\tilde{\phi}}_k' &= \delta\tilde{{\pi}}_k \\
\delta\tilde{{\pi}}_k' &= \left(\frac{1}{2}\tilde{\pi}^2 - 3 \right) \left( \delta\tilde{{\pi}}_k + \left[\tilde{\pi}^2 + \frac{e^{2(\ln{\tilde{k}} - N)}}{\tilde{V}} + \frac{2 \tilde{\pi} \tilde{V}_{,\phi}}{\tilde{V}} + \frac{\tilde{V}_{,\phi\phi}}{\tilde{V}} \right] \delta{\tilde{\phi}}_k \right) \equiv g_k(N, \tilde{\phi}, \tilde{\pi}, \delta{\tilde{\phi}}_k, \delta\tilde{{\pi}}_k) \ ,
\end{align}
\begingroup
where prime denotes derivative with respect to the number of e-folds, and we have introduced the dimensionless variables $\tilde{\phi} \equiv \frac{\phi}{M_{\text{Pl}}}$, $\tilde{\pi} \equiv \frac{\pi}{M_{\text{Pl}}}$, $\delta{\tilde{\phi}}_k \equiv \sqrt{2 H_0 \tilde{k}^3} \delta{\phi}_k$, $\delta\tilde{{\pi}}_k \equiv \sqrt{2 H_0 \tilde{k}^3} \delta{\pi}_k$, $\tilde{V} \equiv \frac{V}{H_0^2 M_{\text{Pl}}^2}$, $\tilde{V}_{,\phi} \equiv \frac{V_{,\phi}}{H_0^2 M_{\text{Pl}}}$, $\tilde{V}_{,\phi\phi} \equiv \frac{V_{,\phi\phi}}{H_0^2}$, and $\tilde{k} \equiv \frac{k}{H_0}$, with $H_0$ being the initial value of the Hubble rate in the beginning of the simulation.

To numerically integrate our equations of motion, we use an explicit Runge--Kutta method of order 4 with fixed time step $\dd{N}$. Denoting consecutive time steps by $i$, the discretised solutions are
\begin{align}
\tilde{\phi}_{i+1} &= \tilde{\phi}_i + \dd{N}\tilde{\pi}_i + \frac{1}{6}\dd{N}^2 (m_1 + m_2 + m_3) \\
\tilde{\pi}_{i+1} &= \tilde{\pi}_i + \frac{1}{6}
\dd{N} (m_1 + 2 m_2 + 2 m_3 + m_4) \\
\delta{\tilde{\phi}}_{k,i+1} &= \delta{\tilde{\phi}}_{k,i} + \dd{N} \delta\tilde{{\pi}}_{k,i} + \frac{1}{6}\dd{N}^2 (n_{k,1} + n_{k,2} + n_{k,3}) \\
\delta\tilde{{\pi}}_{k,i+1} &= \delta\tilde{{\pi}}_{k,i} + \frac{1}{6}\dd{N} (n_{k,1} + 2 n_{k,2} + 2 n_{k,3} + n_{k,4}) \ , 
\end{align}
where
\begin{align}
m_1 &= f(\tilde{\phi}_i, \tilde{\pi}_i) \\
m_2 &= f(\tilde{\phi}_i + \tfrac{1}{2}\dd{N}\tilde{\pi}_i, \tilde{\pi}_i + \tfrac{1}{2}\dd{N} m_1) \\
m_3 &= f(\tilde{\phi}_i + \tfrac{1}{2}\dd{N}\tilde{\pi}_i + \tfrac{1}{4}\dd{N}^2 m_1, \tilde{\pi}_i + \tfrac{1}{2}\dd{N} m_2) \\
m_4 &= f(\tilde{\phi}_i + \dd{N}\tilde{\pi}_i + \tfrac{1}{2}\dd{N}^2 m_2, \tilde{\pi}_i + \dd{N} m_3) \\
n_{k,1} &= g_k(N_i, \tilde{\phi}_i, \tilde{\pi}_i, \delta{\tilde{\phi}}_{k,i}, \delta\tilde{{\pi}}_{k,i}) \\
n_{k,2} &= g_k(N_i + \tfrac{1}{2}\dd{N}, \tilde{\phi}_i + \tfrac{1}{2}\dd{N}\tilde{\pi}_i, \tilde{\pi}_i + \tfrac{1}{2}\dd{N} m_1, \nonumber \\
&\qquad\qquad \delta{\tilde{\phi}}_{k,i} + \tfrac{1}{2}\dd{N}\delta\tilde{{\pi}}_{k,i}, \delta\tilde{{\pi}}_{k,i} + \tfrac{1}{2}\dd{N} n_{k,1}) \\
n_{k,3} &= g_k(N_i + \tfrac{1}{2}\dd{N}, \tilde{\phi}_i + \tfrac{1}{2}\dd{N}\tilde{\pi}_i + \tfrac{1}{4}\dd{N}^2 m_1, \tilde{\pi}_i + \tfrac{1}{2}\dd{N} m_2, \nonumber \\
&\qquad\qquad \delta{\tilde{\phi}}_{k,i} + \tfrac{1}{2}\dd{N}\delta\tilde{{\pi}}_{k,i} + \tfrac{1}{4}\dd{N}^2 n_{k,1}, \delta\tilde{{\pi}}_{k,i} + \tfrac{1}{2}\dd{N} n_{k,2}) \\
n_{k,4} &= g_k(N_i + \dd{N}, \tilde{\phi}_i + \dd{N}\tilde{\pi}_i + \tfrac{1}{2}\dd{N}^2 m_2, \tilde{\pi}_i + \dd{N} m_3, \nonumber \\
&\qquad\qquad \delta{\tilde{\phi}}_{k,i} + \dd{N}\delta\tilde{{\pi}}_{k,i} + \tfrac{1}{2}\dd{N}^2 n_{k,2}, \delta\tilde{{\pi}}_{k,i} + \dd{N} n_{k,3}) \ ,
\end{align}
and the real and imaginary parts of the perturbation modes are solved independently.

In the full treatment of the kicks, the stochastic field kicks $\Delta{\tilde{\phi}_i}$ are normally distributed with variance
\begin{align}
\expval{(\Delta\tilde{\phi}_i)^2} &= \eval{\zeta\dd{N} \left(1 - \frac{1}{2}\tilde{\pi}_i^2 \right) \abs{\delta{\tilde{\phi}}_{k,i}}^2}_{k=\sigma a H} \ , 
\end{align}
where $\zeta \equiv \frac{H_0^2}{4 \pi^2 M_{\text{Pl}}^2}$, and the subscript $k = \sigma a H$ indicates that the mode amplitude is taken at the coarse-graining scale. The momentum kicks are proportional to the realised field kicks,
\begin{align}
\Delta{\tilde{\pi}_i} = \eval{\Delta{\tilde{\phi}_i} \frac{\Re(\delta\tilde{{\pi}}_{k,i} \delta{\tilde{\phi}}_{k,i}^*)}{\abs{\delta{\tilde{\phi}}_{k,i}}^2}}_{k=\sigma a H} \ .
\end{align}
The proportionality factor is equal to $\Re (\delta\tilde{{\pi}}_{k,i}/\delta\tilde{{\phi}}_{k,i})$ discussed below \eqref{eq:squeezed_p}, just written in a more convenient form.

In the recursive treatment, the mode amplitudes in the above expressions are replaced with ones pre-calculated on a non-stochastic FLRW background, so that no perturbation modes are evolved during the simulation. In the simplified treatment, complexity is reduced even further by removing also the pre-calculated modes, setting the field kick variance to
\begin{align}
\expval{(\Delta\tilde{\phi}_i)^2} &= \zeta\dd{N} \frac{\tilde{V}}{3 - \frac{1}{2}\tilde{\pi}_i^2} \ ,
\end{align}
and removing the momentum kicks entirely.

For the perturbations, we consider a discrete grid of modes with modulus evenly distributed on a logarithmic scale as $\ln(k_{i+1}) = \ln(k_i) + \dd{k}$, where $\dd{k}$ is the mode spacing. The evolution of each mode begins when $k = \alpha a H$, with $\alpha = 100$ (the results are insensitive to making $\alpha$ larger). The initial conditions are $\delta{\tilde{\phi}}_k = \alpha\tilde{H}$ and $\delta{\tilde{\pi}}_k = -\alpha\tilde{H} - i \alpha^2 \tilde{H}$, where $\tilde{H}\equiv\frac{H}{H_0}$. The longest wavelength mode we consider corresponds to the CMB pivot scale $k_*$, and its evolution starts immediately at the onset of each simulation. The next mode to begin evolving is denoted by $k_\mr{next}$. For each realisation, the code executes the algorithm below:\\

\begin{algorithm}[H]
\SetAlgoVlined
\DontPrintSemicolon
\SetInd{0.5em}{1em}
\SetAlgoHangIndent{1em}
\SetVlineSkip{0.5em}
Set initial values for $N$, $\phil$, $\pil$. Set $k_\mr{next} = k_\mr{*}$. Set current kick coefficient to zero.\;
\While{$\phil > \phif$}
{
    Evolve $N$, $\phil$, $\pil$.\;
    \For{all modes k in the simulation}
    {
        \eIf{$k > \sigma a H$}
        {
            Evolve $\delta\phi_{\vec{k}}$, $\delta\phi_{\vec{k}}'$.
        }
        {
            Evolve $\delta\phi_{\vec{k}}$, $\delta\phi_{\vec{k}}'$ to $k = \sigma a H$. Set the current kick coefficient from $\delta\phi_{\vec{k}}$, $\delta\phi_{\vec{k}}'$. Remove mode $k$ from the simulation.
        }
    }
    \eIf{$k_\mr{next} \leq \kpbh$}
    {
        \If{$k_\mr{next} \leq \alpha a H$}
        {
            Add mode $k = k_\mr{next}$ to the simulation. Set initial values for $\delta\phi_{\vec{k}}$, $\delta\phi_{\vec{k}}'$. Evolve $\delta\phi_{\vec{k}}$, $\delta\phi_{\vec{k}}'$ from $k = \alpha a H$. Set $k_\mr{next} = e^{\dd{k}} k_\mr{next}$.
        }  
    }
    {
        \If{$k_\mr{next} \leq \sigma a H$}
        {
            Set the current kick coefficient to zero.
        }   
    }

    \If{$a H \geq \max(a H)$}
    {
      Add stochastic kick to $\phil$, $\pil$ using the current kick coefficient.
    }
}
\caption{Evolution for each run}
\end{algorithm}\vspace*{0.5cm} 

Initially, the background field is in slow-roll. Although our code generally runs on a fixed time step, it takes partial steps for increased accuracy whenever we encounter events that are not synchronised with the time steps, \eg when we start or stop evolving modes, and at the end of inflation. The condition $a H \geq\max(a H)$ guarantees that we do not apply duplicate kicks in the event that $a H$ momentarily decreases in the middle of the simulation, as happens for the relic potential.

The values used for the time step $\dd{N}$ and the mode spacing $\dd{k}$ depend on the potential. The time steps are generally between $1/256$ and $1/512$, and the mode spacings are between $1/32$ and $1/256$. These values are the result of extensive testing for each of the potentials; they have been chosen so that any further halving of the time step or mode spacing would change the results by less than 1\%.

\section{Gauge transformations} \label{sec:gauge}

Under a gauge transformation from the spatially flat gauge to the uniform $N$ gauge, the field perturbation and its derivative transform as
\begin{align}
\delta{\phi}_k &\rightarrow \delta{\phi}_k + \phi' \alpha_k \\
\delta{\phi}_k' &\rightarrow \delta{\phi}_k' + (\phi' \alpha_k)' \ ,
\end{align}
where $\alpha_k$ is a gauge transformation parameter defined in \cite{Pattison:2019hef}, and in this appendix prime exceptionally denotes a derivative with respect to conformal time. We use the same initial condition as in \cite{Pattison:2019hef}, $\alpha_k=0$.

Figure \ref{fig:alpha} shows the amplitude of the gauge correction terms relative to the corresponding perturbation at the coarse-graining scale. In the asteroid, solar and supermassive cases, the relative correction is less than $10^{-5}$ for most modes. However, there is a single spike in each plot just before or during USR. For these modes the amplitude of the perturbation (or its derivative) in the denominator becomes small, driving up the value of the fraction. The contribution of such modes to the overall dynamics is very small. For the relic case the relative gauge corrections are much larger, reaching almost order unity in general, and even larger for the spikes. Note that we show the modes at the coarse-graining scale, not at Hubble exit. In the relic case, there are of order unity gauge corrections during USR even at Hubble crossing. In \cite{Pattison:2019hef} it was assumed that in USR the potential can be neglected completely, and that $\sigma\ll1$. These approximations do not hold in our case.

\newpage

\begin{figure}[H]
\centering
\includegraphics{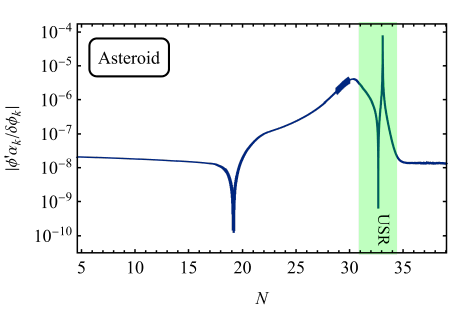}
\includegraphics{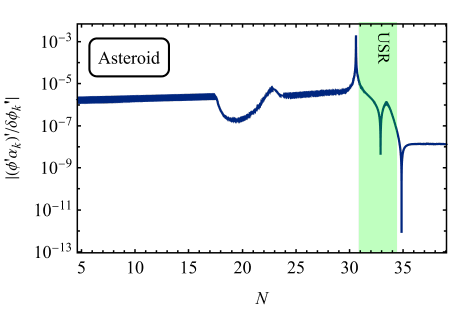}
\includegraphics{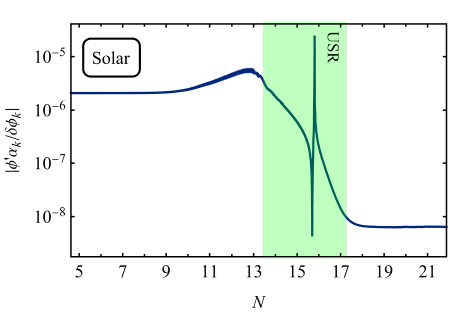}
\includegraphics{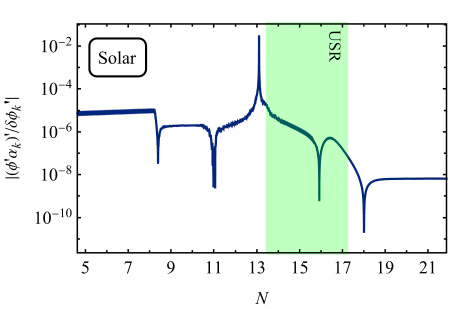}
\includegraphics{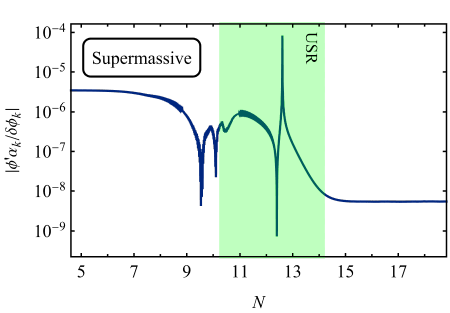}
\includegraphics{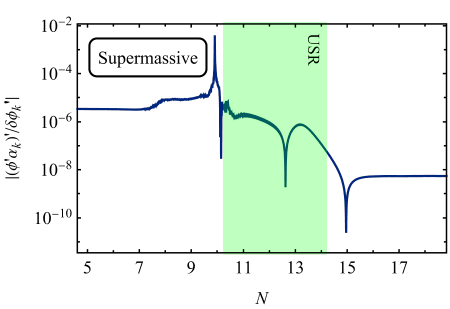}
\includegraphics{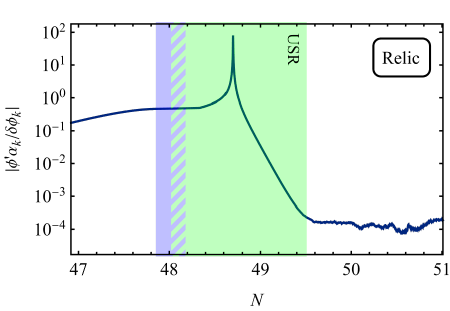}
\includegraphics{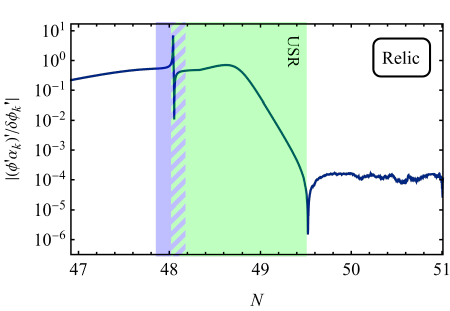}
\caption{Relative gauge transformation corrections to $\delta{\phi}_k$ (left column) and $\delta{\phi}_k'$ (right column) at the coarse-graining scale for the asteroid, solar, supermassive and relic cases, in order from top to bottom. The green area indicates the USR region, the violet area indicates the period where inflation temporarily ends, and its overlap with the USR region is marked by the diagonal lines.}
\label{fig:alpha}
\end{figure}

\bibliographystyle{JHEP}
\bibliography{stoc}

\end{document}